\documentclass[trackchanges,twocolumn]{aastex7}
\usepackage{comment}
\usepackage{mathrsfs}
\usepackage{graphicx}
\usepackage{subfigure}

\usepackage{tablefootnote} 
\usepackage{multirow}
\usepackage{makecell} 
\usepackage{colortbl} 
\usepackage[table]{xcolor} 

\usepackage{bm}

\newcommand{\f}[2]{\frac{#1}{#2}}
\newcommand{\mr}[1]{\mathrm{#1}}
\newcommand{\p}{\partial}
\newcommand{\ten}[1]{\bm{\mathsf{#1}}}
\newcommand{\h}{\hspace{1mm}}
\newcommand{\bra}[1]{\left(#1\right)}
\newcommand{\ctext}[1]{\raise0.2ex\hbox{\textcircled{\scriptsize{#1}}}}
\newcommand{\chH}{\mathrm{H}}
\newcommand{\chHp}{\mathrm{H}^{+}}
\newcommand{\chHm}{\mathrm{H}^{-}}
\newcommand{\chHt}{\mathrm{H}_{2}}

\newcommand{\eqref}[1]{(\ref{#1})}
\newcommand{\fdis}{\eta}
\newcommand{\chirad}{\xi}
\newcommand{\EofRSLA}{E^\mr{R}}
\newcommand{\EofRSLAijk}{E^{\mr{R},i,j,k}}
\newcommand{\EofRSLARorL}{E^{\mr{R},\mr{R\,or\,L}}}
\newcommand{\NofRSLA}{N^\mr{R}}
\newcommand{\PofRSLA}{\ten{P}^\mr{R}}
\newcommand{\tenQofRSLA}{\ten{Q}^\mr{R}}
\newcommand{\bmGofRSLA}{\bm{G}^\mr{R}}
\newcommand{\bmGofRSLAprime}{\bm{G}^{\prime \mr{R}}}
\newcommand{\EofNonRSLA}{E^\mr{N}}
\newcommand{\EofNonRSLAijk}{E^{\mr{N},i,j,k}}
\newcommand{\EofNonRSLAR}{E^{\mr{N},\mr{R}}}
\newcommand{\EofNonRSLAL}{E^{\mr{N},\mr{L}}}
\newcommand{\EofNonRSLARorL}{E^{\mr{N},\mr{R\,or\,L}}}
\newcommand{\NofNonRSLA}{N^\mr{N}}
\newcommand{\FofNonRSLA}{F^\mr{N}}
\newcommand{\bmFofNonRSLA}{\bm{F}^\mr{N}}
\newcommand{\bmFofNonRSLAn}{\bm{F}^{\mr{N},n}} 
\newcommand{\bmGofNonRSLA}{\bm{G}^\mr{N}}

\newcommand{\fdisofNonRSLA}{\fdis^\mr{N}} 

\begin{document}

\title{An Explicit M1 Radiation-hydrodynamics Scheme for 3D Protostellar Evolution}

\author[orcid=0000-0001-8382-3966]{Kazutaka Kimura}
\affiliation{Astronomical Institute, Graduate School of Science, Tohoku University, Aoba, Sendai 980-8578, Japan}
\email[show]{kimura.k@astr.tohoku.ac.jp}  

\author[orcid=0000-0001-7842-5488]{Kazuyuki Sugimura} 
\affiliation{Faculty of Science, Hokkaido University, Sapporo, Hokkaido 060-0810, Japan}
\email{fakeemail2@google.com}

\author[orcid=0000-0003-3127-5982]{Takashi Hosokawa} 
\affiliation{Department of Physics, Graduate School of Science, Kyoto University, Sakyo, Kyoto 606-8502, Japan}
\email{fakeemail2@google.com}

\author[orcid=0000-0002-0547-3208]{Hajime Fukushima} 
\affiliation{Center for Computational Sciences, University of Tsukuba, Ten-nodai, 1-1-1 Tsukuba, Ibaraki 305-8577, Japan}
\email{fakeemail2@google.com}

\author[orcid=0000-0001-5922-180X]{Kazuyuki Omukai}
\affiliation{Astronomical Institute, Graduate School of Science, Tohoku University, Aoba, Sendai 980-8578, Japan}
\email{fakeemail3@google.com}

\begin{abstract}
We present a radiation-hydrodynamics (RHD) scheme that enables 3D simulations resolving both protostellar interiors and their surrounding accretion flows within a single framework, to clarify how a protostar evolves while interacting with the accretion flow.
The method builds on an explicit two-moment M1 closure scheme with a reduced speed of light approximation (RSLA) for massively parallel computation.
Our scheme introduces a complementary non-RSLA radiation component that dominates in optically thick regions. 
This hybrid treatment restores physical energy conservation inside protostars, which would otherwise be violated under the RSLA, while retaining the advantage of large time steps.
To overcome the limitation of the conventional M1 closure in solving radiative transfer in extremely optically thick regions inside protostars and across steep optical-depth gradients near their surfaces, we incorporate the optical-depth information of neighboring cells into the radiative transfer calculation.
We further evolve photon-number densities in addition to radiation energy densities to reconstruct an effective local spectrum on the fly without resorting to costly multifrequency transport. 
We implement this scheme in the adaptive mesh refinement code SFUMATO and verify its validity through a series of test calculations. 
As an application, we follow the early evolution of a massive protostar formed at high redshift, within a full cosmological context.
The results reveal a continuous structure connecting the swollen protostar and its surrounding disk, which cannot be captured in conventional 1D models.
This RHD scheme opens a path to studies of protostellar evolution and its interaction with the accretion flow in realistic 3D environments.
\end{abstract}



\section{INTRODUCTION}
The protostellar evolution with mass accretion represents the fundamental stage during which a star gains most of its mass. The nature of this process determines the final stellar masses and, in turn, the shape of the stellar mass spectrum across cosmic time, both in the early Universe and in present-day star-forming environments \citep[see recent reviews by][]{Klessen_and_Glover2023, Hennebelle_and_Grudic_2024, Beuther_et_al_2025}.
\par
Protostars evolve through complex interactions with the accreting gas surrounding them. The accretion flow supplies entropy and angular momentum to the protostars, thereby influencing their internal structure and evolutionary path. Conversely, massive protostars exert strong radiative and mechanical feedback on the surrounding gas, regulating the subsequent accretion process.
In recent studies, large-scale accretion dynamics under the protostellar feedback have primarily been investigated with 3D radiation-hydrodynamic (RHD) simulations \citep[e.g.,][]{Krumholz_et_al_2007_Simulations,Kuiper_et_al_2011, Susa_2013,Hosokawa_et_al_2016,Klassen_et_al_2016,Rosen_et_al_2016,Stacy_et_al_2016,Chon_et_al_2018,Chon_et_al_2024,Sugimura_et_al_2020,Sugimura_et_al_2023,Jaura+2022,Sharda_and_Menon_2025}. 
In such simulations, however, the protostar itself is typically treated with subgrid models, such as sink particles, and neither its internal structure nor the boundary layer between the protostar and the accretion disk is resolved explicitly.
Meanwhile, the structure and evolution of protostars on sub-astronomical unit scales have been studied with 1D stellar evolution calculations \citep[e.g.,][]{Stahler_et_al_1980,Stahler_et_al_1986,Palla_and_Stahler_1991, Behrend+Maeder01, Omukai_and_Palla_2001,Omukai_and_Palla_2003,Yorke_and_Bodenheimer08,Hosokawa_and_Omukai_2009b, Hosokawa_et_al_2010, Hosokawa_et_al_2013,Sakurai_et_al_2015,Haemmerle_et_al_2018_evol,Haemmerle_et_al_2018_rot,Herrington_et_al_23,Nandal_et_al_23, Nandal_et_al_24}. 
These 1D models typically assume spherically symmetric accretion onto the protostar, a hydrostatic structure within the stellar interior, and a simple rotational motion of the protostar itself, thereby neglecting both the inherently anisotropic nature of gas inflow and the resulting complex dynamics within the protostar.
This separation of scales has limited our ability to fully capture the mutual influence between the protostar and its accreting environment.
\par
To overcome this limitation, it is essential to resolve both the stellar interior and the large-scale accretion flow in a unified 3D RHD simulation. 
Recent developments in computational resources and numerical schemes have made such simulations increasingly feasible \citep{Luo_et_al_2018,Bhandare_et_al_2020,Bhandare_et_al_2024,Bhandare_et_al_2025,Ahmad_et_al_2023,Ahmad_et_al_2024,Kimura_et_al_2023,Mayer_et_al_2025_arXiv}.
A widely adopted radiative transfer (RT) method in these simulations is the flux-limited diffusion (FLD) approximation with implicit time integration.
The FLD method reduces the computational cost by replacing the full RT problem, which involves three spatial dimensions, two angular dimensions, and one frequency dimension, with a single zeroth moment diffusion equation \citep{Levermore_and_Pomraning_1981}.
Moreover, the implicit time integration enables us to take time steps longer than the timescale required for stable explicit integration.
This is particularly advantageous in star formation simulations, where the dynamical timescale of the gas is much longer than the RT timescale.
\par
In addition to the FLD method with implicit time integration, the two-moment M1 closure method combined with the reduced speed of light approximation (RSLA) has also been widely employed in modern star formation RHD simulations \citep{Levermore_1984,Skinner_and_Ostriker_2013,Rosdahl_and_Teyssier_2015,Kannan_et_al_2019,Mignon-Risse_et_al_2020,Fukushima_and_Yajima_2021,Yaghoobi_et_al_2025}.
The M1 closure scheme solves not only the zeroth but also the first moment of the RT equation, which allows for a more accurate treatment of anisotropic radiation fields and prevents unphysical diffusion compared to the FLD.
In addition, the RSLA alleviates the severe time step constraints imposed by the Courant–Friedrichs–Lewy (CFL) condition in M1 closure, enabling stable explicit integration with large time steps \citep{Gnedin_and_Abel_2001}.
More sophisticated approaches, such as the Variable Eddington Tensor method, have been used to overcome the limitations inherent to the M1 closure scheme, for instance its inability to accurately treat multibeam radiation fields \citep[e.g.,][]{Stone_et_al_1992,Sharda_and_Menon_2025}.
However, owing to the balance between accuracy and computational cost, the M1 closure scheme still remains widely adopted in current large-scale star formation simulations.
\par
Such an explicit formulation is highly suitable for massively parallel computing and GPU-accelerated architectures, and can outperform FLD with implicit time integration depending on the simulation setup \citep{Wibking_et_al_2022}.
Therefore, to follow realistic protostellar evolution over longer timescales in massively parallel simulations, it is necessary to apply the M1 closure scheme with the RSLA to simulations that resolve both the protostellar interior and the large-scale accretion flow simultaneously.
However, direct application of the M1 closure scheme with the RSLA gives rise to the following challenges.
\par
The first challenge is ensuring total energy conservation in protostars.
When the RSLA is employed, the rate of energy exchange between gas and radiation is artificially reduced to reproduce the steadystate of the system with the full speed of light.
However, this modification causes the loss of physical energy during the energy exchange, thereby leading to violation of total energy conservation \citep{Wibking_et_al_2022}.
This issue is not problematic in optically thin regions outside protostars, where energy naturally escapes from the system as radiation.
In optically thick environments, such as the interiors of protostars, however, this artificial energy loss effectively removes entropy from the system, preventing accurate modeling of protostellar evolution.
\par
The second challenge concerns the treatment of RT within the protostellar interior and across its surface.
In star formation RHD simulations, the computational grid is typically refined according to the Jeans criterion, i.e., resolving the Jeans length by a prescribed number of cells.
However, even when the Jeans length is resolved with several to a few tens of cells, the optical depth per grid cell in the protostellar interior still far exceeds unity.
In such regions, the M1 closure method introduces excessive numerical diffusion to ensure integration stability. 
To mitigate this issue, \citet{Rosdahl_and_Teyssier_2015} proposed a scheme that splits the radiation energy into two components, which effectively suppresses artificial diffusion and reproduces the physical behavior of radiative diffusion in optically thick interiors \citep[see also][]{He_et_al_2024_RHD}. 
At the protostellar surface, however, steep gradients of optical depth naturally arise, where radiation transitions from being trapped to freely escaping. 
In this regime, their method of splitting the radiation energy into two components prevents radiation from readily leaking across the sharp gradient, causing the scheme to underestimate the outgoing radiation flux despite its accuracy in the interior.
\par
The third challenge concerns the treatment of the radiation energy spectrum, which is a common limitation of previous RHD simulations including those using the FLD method.
Spectral information is essential for accurately evaluating the protostellar radiative feedback on the surrounding gas, including heating and ionization.
However, many RHD simulations assume a fixed spectral shape rather than following its evolution explicitly.
For example, it is common to adopt a spectrum based on simplified 1D protostellar models from previous studies.
However, in simulations resolving the protostar itself that we focus on here, the spectral evolution may deviate significantly from these simplified models. 
Thus, we need to estimate it on the fly in a manner consistent with the actual protostellar evolution in simulations.
The most straightforward approach to determining the radiation spectrum is to divide the energy into multiple frequency bins and solve the emission, absorption, and transport processes for each bin \citep{Yorke_and_Sonnhalter_2002,Kuiper_et_al_2010,Klassen_et_al_2014, Rosen_and_Krumholz_2017,Kuiper_et_al_2018}.
While this method provides high accuracy, it substantially increases computational cost.
Therefore, an alternative method with lower computational cost is required.
\par
In this study, we propose a new RHD scheme that addresses the challenges discussed above.
Our RT approach is based on the explicit M1 closure method with the RSLA, which follows the anisotropy of the radiation field and ensures high parallel efficiency for large-scale computations.
We have implemented the scheme in the self-gravitating magnetohydrodynamics code SFUMATO \citep{Matsumoto_2007,Matsumoto_et_al_2015}, which has been widely applied to various problems of star formation 
\citep[e.g.][]{Inoue_et_al_2018,Matsumoto_et_al_2019,Fukushima_et_al_2020,Sugimura_et_al_2020,Sugimura_et_al_2023, Abe_et_al_2021,Fukushima_and_Yajima_2021,Kimura_et_al_2021, Kimura_et_al_2023,Sadanari_et_al_2021,Sadanari_et_al_2024}.
Through a series of tests, we demonstrate the accuracy and robustness of the method.
Furthermore, we apply this scheme to primordial star formation in the early Universe and show how protostars evolve in conjunction with the surrounding large-scale gas dynamics.
\par
The remainder of this paper is organized as follows.
In Section \ref{sec:Review}, we review the basic RHD formulation of the M1 closure scheme, combined with the multicomponent method proposed by \citet{Rosdahl_and_Teyssier_2015} and the RSLA.
In Section \ref{sec:challenges}, we provide an overview of the three challenges discussed above.
Section \ref{sec:improved_scheme} describes the formulation of our improved RHD scheme to solve the challenges. 
In Section \ref{sec:comp_proc}, we present its numerical implementation in the self-gravitating magnetohydrodynamics code, SFUMATO.
In Section \ref{sec:tests}, we validate our method using a series of test problems.
Section \ref{sec:ProsForm} presents simulation results for the evolution of a protostar formed in the early Universe as an application of our new scheme.
In Section \ref{sec:Discussion}, we discuss the limitation of our new scheme.
Finally, Section \ref{sec:Summary} provides a summary of this paper.
\section{Review of RHD Formulation} \label{sec:Review}
This section presents the RHD formulation of the M1 closure scheme with the multicomponent method proposed by \citet{Rosdahl_and_Teyssier_2015} and the RSLA. 
In Section~\ref{subsec:RHD_eq}, we introduce the basic RHD equations for star formation simulations, where the gas velocity is much smaller than the speed of light.
In Section~\ref{subsec:RT15}, we describe the method by \citet{Rosdahl_and_Teyssier_2015}, which modifies the M1 closure scheme to follow RT accurately in highly optically thick regions.
In Section~\ref{subsec:RSLA}, we review the RSLA formulation, which relaxes the CFL constraint imposed by the speed of light and allows for larger time steps.
\subsection{Basic RHD equations with M1 closure} \label{subsec:RHD_eq}
Here, we introduce the RHD equations.
Throughout this paper, we restrict our discussion to the regime where $\beta = |\bm{v}| / c \ll 1$, with $\bm{v}$ denoting the gas velocity and $c$ the speed of light.
We take into account both absorption and scattering processes, including their frequency dependence.
The formulation of the RHD equations depends on the choice of reference frame, specifically, whether physical quantities are expressed in the lab frame or in the comoving-frame.
In this study, we adopt the so-called mixed frame formulation, in which opacities and emissivities are evaluated in the comoving frame, while all other gas and radiation quantities are defined in the lab frame \citep{Mihalas_and_Klein_1982,Krumholz_et_al_2007_Algorithms}.
This formulation enables the numerical algorithm to conserve total energy to machine precision during gas–radiation interaction. 
Moreover, because radiation energy in the lab frame is a conserved quantity, it is obvious how to refine or coarsen the grid in a conservative manner in AMR codes. 
In this paper, physical quantities measured in the comoving frame are denoted with the subscript $0$.
We show the Lorentz transformations between the lab frame and comoving frame quantities in Appendix~\ref{sec:app_LT}.
\par
In the regime where $\beta \ll 1$, the behavior of the radiation–gas interaction depends not only on the velocity ratio $\beta$ but also on the optical depth of the system $\tau$.
There exist three characteristic limiting regimes:
the optically thin limit ($\tau \ll 1$), where radiation propagates freely;
the static diffusion limit ($\tau \gg 1$, $\beta \tau \ll 1$), where radiation diffuses through a nearly stationary medium; and
the dynamic diffusion limit ($\tau \gg 1$, $\beta \tau \gg 1$), where radiation is advected along with the gas flow.
The relative importance of terms in the RHD equations varies across these regimes.
Therefore, to retain only the leading-order terms in the limit of $\beta \ll 1$, it is necessary to evaluate which terms dominate in each limiting case.
In Appendix~\ref{sec:app_dom}, we derive the governing RHD equations by retaining all the dominant terms in each of these limits, and summarize how these equations relate to the formulations presented in previous studies.
The derived equations are
\begin{eqnarray}
    && \f{\p \rho}{\p t} + \nabla\cdot(\rho\bm{v}) = 0 , \label{eq_basic:rho} \\
    && \f{\p E_\mr{gas}}{\p t} + \nabla\cdot\left\{\bm{v}(E_\mr{gas}+P_\mr{gas})\right\} = \rho\bm{g}\cdot\bm{v} + c G^0  ,\label{eq_basic:Egas} \\
    && \f{\p \rho\bm{v}}{\p t} + \nabla\cdot(\rho\bm{v}\otimes\bm{v}+P_\mr{gas}\ten{I}) = 
    \rho\bm{g} + \bm{G} , \label{eq_basic:rhov} \\
    && \f{\p E_\mr{rad}}{\p t} + \nabla\cdot\bm{F} = -cG^0 , \label{eq_basic:Erad} \\
    && \f{\p \bm{F}}{\p t} + c^2\nabla\cdot\ten{P}_\mr{rad} = -c^2 \bm{G} , \label{eq_basic:F} \\
    && G^0 = \rho ( \kappa_{0E} E_\mr{rad}
    - \kappa_\mr{0P} a T^4) , \label{eq_basic:G0} \\
    && \bm{G} = \rho\chi_{0F}\f{\bm{F}}{c}-\rho\chi_{0F}\f{4\bm{v}}{3c}E_\mr{rad}, \label{eq_basic:G}
\end{eqnarray}
where $\ten{I}$ denotes the identity tensor, $a$ is the radiation constant, $\bm{g}$ is the gravitational acceleration, and $\rho$, $P_\mathrm{gas}$ and $T$ are the gas density, pressure, and temperature, respectively. We assume the equation of state $P_\mr{gas} = k_\mr{B} T \rho/\mu m_\mr{H}$, where $k_\mr{B}$ is the Boltzmann constant, $m_\mr{H}$ is the mass of a hydrogen nucleus, and $\mu$ is the mean molecular weight.
The total energy density of the gas $E_\mathrm{gas}$ is given by $\rho \bm{v}^2/2 + P_\mr{gas}/(\gamma-1)$, where $\gamma$ is the specific heat ratio. 
The quantities $E_\mathrm{rad}$, $\bm{F}$, and $\ten{P}_\mathrm{rad}$ denote the radiation energy density, radiation flux, and radiation-pressure tensor in the lab frame, respectively.
The 4D vector $(G^0,\bm{G})$ is the radiation four-force density representing the gas-radiation interaction through emission, absorption, and scattering.
Note that, in simulations that do not account for the dynamic diffusion regime, we can ignore the second term on the right-hand side of Equation~\eqref{eq_basic:G}.
\par
The three opacities appearing in $(G^0,\bm{G})$ given by Equations~\eqref{eq_basic:G0} and \eqref{eq_basic:G} are the Planck mean, energy-mean, and flux-mean opacities, which are defined as
\begin{eqnarray}
    && \kappa_\mr{0P} = \f{\int_0^\infty d\nu \h \kappa^\nu_{0} B^\nu(T)}{\int_0^\infty d\nu \h B^\nu(T)} , \label{eq:kappa_0P} \\
    && \kappa_{0E} = \f{\int_0^\infty d\nu \h \kappa^\nu_{0} E^\nu_{\mr{rad},0}}{\int_0^\infty d\nu \h E^\nu_{\mr{rad},0}} , \label{eq:kappa_0E} \\
    && \chi_{0F} = \f{\int_0^\infty d\nu \h (\kappa^\nu_{0} + \sigma^\nu_{0}) F^\nu_{0}}{\int_0^\infty d\nu \h F^\nu_{0}} , \label{eq:kappa_0F}
\end{eqnarray}
where $\kappa^\nu_{0}$ and $\sigma^\nu_{0}$ are the frequency-dependent absorption and scattering opacities evaluated in the comoving-frame, $B^\nu(T)$ is the Planck distribution at temperature $T$, and $E^\nu_{\mr{rad},0}$ and $F^\nu_{0}$ are the comoving frame radiation energy density and flux per unit frequency.
For calculating $\kappa_{0E}$ and $\chi_{0F}$ with Equations~\eqref{eq:kappa_0E} and \eqref{eq:kappa_0F}, we need the spectrum of radiation energy $E^\nu_{\mr{rad},0}$ and flux $F^\nu_{0}$.
We describe how to calculate this spectrum with finite frequency bins and low computational cost in Section~\ref{sec:spectral_reconstruction}.
\par
In the M1 closure method, we determine the radiation-pressure tensor  $\ten{P}_\mathrm{rad}$ by imposing the M1 closure relation. 
Specifically, $\ten{P}_\mathrm{rad}$ is expressed in terms of the Eddington tensor $\ten{D}$ as follows \citep{Levermore_1984}:
\begin{eqnarray}
    && \ten{P}_\mr{rad} = \ten{D}E_\mr{rad} , \nonumber \\
    && \ten{D} = \f{1-\chirad}{2}\ten{I} + \f{3\chirad-1}{2}\bm{n}\otimes\bm{n}, \label{eq:D}\\
    && \bm{n}=\f{\bm{F}}{|\bm{F}|}, \h
    \chirad=\f{3+4(|\bm{F}|/(cE_\mr{rad}))^2}{5+2\sqrt{4-3(|\bm{F}|/(cE_\mr{rad}))^2}}
    . \nonumber 
\end{eqnarray}
With these definitions, the M1 closure reproduces the correct physical behavior of radiation in both the optically thin and optically thick limits.
In the optically thin limit where $|\bm{F}|/(cE_\mr{rad}) \simeq 1$  (or $\chirad \simeq 1$), the Eddington tensor $\ten{D}$ approaches $\bm{n} \otimes \bm{n}$, corresponding to a beam-like radiation field.
In the optically thick limit where $|F|/(cE_\mr{rad}) \simeq 0$ (or $\chirad \simeq 1/3$), it approaches $\ten{I}/3$, consistent with an isotropic radiation field.
\subsection{Streaming/trapped component for an optically thick medium} \label{subsec:RT15}
The simple M1 closure method presented above is unable to accurately capture radiative diffusion when the system is optically thick and the optical depth per computational cell, $\tau_\mathrm{cell} = \rho \chi_\mathrm{0R} \Delta x$, exceeds unity.
Here, $\Delta x$ is the computational cell width, and $\chi_\mr{0R}$ is the Rosseland mean opacity, which is defined as
\begin{eqnarray}
    \f{1}{\chi_\mr{0R}} = \f{\int_0^\infty d\nu \h (\kappa^\nu_{0}+\sigma^\nu_{0})^{-1} (\partial B^\nu(T)/\partial T)}{\int_0^\infty d\nu \h (\partial B^\nu(T)/\partial T)}. \label{eq:Rosseland}
\end{eqnarray}
This limitation arises because numerical diffusion, which is introduced to stabilize the numerical integration of the advection terms on the left-hand sides of Equations~\eqref{eq_basic:Erad} and \eqref{eq_basic:F}, can exceed the magnitude of the physical diffusion.
In the optically thick limit, the expected physical diffusion flux is given by
\begin{eqnarray}
    \bm{F}_\mr{diff} = -\f{c}{3\chi_\mr{0R}\rho}\nabla E_\mr{rad}, \label{eq:dif_app_E}  
\end{eqnarray}
which decreases as the optical depth increases.
By contrast, the magnitude of the numerical diffusion is independent of the optical depth.
For example, the global Lax-Friedrichs (GLF) Riemann solver, which is used in this study for its simplicity and its ability to preserve symmetry more robustly than higher order schemes \citep{Rosdahl_et_al_2013}, yields a numerical diffusion of the form
\begin{eqnarray}
    F_\mr{num} = -\f{c}{2}(E_\mr{rad}^\mr{R} - E_\mr{rad}^\mr{L}) , 
\end{eqnarray}
where $E_\mr{rad}^{R}$ and $E_\mr{rad}^\mr{L}$ denote $E_\mr{rad}$ in the right and left neighboring cells, respectively.
This expression can be formally rewritten as
\begin{eqnarray}
    \bm{F}_\mr{num} = - \f{c \Delta x}{2}\nabla E_\mr{rad}. \label{eq:num_diff_E}  
\end{eqnarray}
A comparison between Equations~\eqref{eq:dif_app_E} and \eqref{eq:num_diff_E} reveals that, in regions where $\tau_\mathrm{cell} > 2/3$, the numerical diffusion $\bm{F}_\mr{num}$ exceeds the physical diffusion $\bm{F}_\mr{diff}$, thereby limiting the accuracy of the M1 scheme.
\par
To address this issue, \citet{Rosdahl_and_Teyssier_2015} proposed a method that reproduces the correct physical diffusion in the optically thick limit.
Since the magnitude of the numerical diffusion scales with $E_\mr{rad}$, as shown in Equation~\eqref{eq:num_diff_E}, one can adjust $\bm{F}_\mathrm{num}$ to match the physical diffusion $\bm{F}_\mathrm{diff}$ by reducing $E_\mr{rad}$ in the RT calculation.
In this approach, they split $E_\mathrm{rad}$ into two components, \textit{streaming photons} $E_\mathrm{rad,S}$ and \textit{trapped photons} $E_\mathrm{rad,T}$, and then use only $E_\mathrm{rad,S}$ for solving the advection terms.
For the GLF solver, these components should be defined as
\begin{equation}
\begin{array}{l}
    E_\mr{rad,S} = (1-\fdis_\mr{T}) E_\mr{rad} , \hspace{3mm} E_\mr{rad,T} = \fdis_\mr{T} E_\mr{rad} , \\ [6pt]
    {\displaystyle \fdis_\mr{T} = \exp\left(-\frac{2}{3\tau_\mr{cell}}\right) .}
\end{array}
\label{eq:distribute_ST}
\end{equation}
Using this definition of $E_\mathrm{rad,S}$, we can obtain the numerical diffusion in the optically thick limit where $\tau_\mr{cell} \gg 1$ by
\begin{eqnarray}
    \bm{F}_\mathrm{num} = - \frac{c\Delta x}{2}\nabla E_\mathrm{rad,S} \simeq - \frac{c}{3\chi_\mathrm{0R}\rho}\nabla E_\mathrm{rad}, \label{eq:Fnum}
\end{eqnarray}
thus recovering the correct physical diffusion described by Equation~\eqref{eq:dif_app_E}.
\par
While using only the streaming photons for solving the advection terms enables the scheme to reproduce the correct physical diffusion, it introduces a new issue.
The radiative flux $\bm{F}$, obtained from Equation~\eqref{eq_basic:F}, now corresponds only to the streaming component, denoted $\bm{F}_\mathrm{S}$, and no longer represents the full radiative flux in regions where $\tau_\mathrm{cell} > 1$ and the trapped component $E_\mathrm{rad,T}$ dominates.
\par
Thus, we must modify the terms involving the radiation flux $\bm{F}$ in the RHD equations to include the flux contribution of $E_\mathrm{rad,T}$ as follows.
The following treatment is essentially equivalent to that described in Appendix~B of \citet{Rosdahl_and_Teyssier_2015}, but without the impact of the RSLA, which we explicitly introduce later in Section~\ref{subsec:RSLA}.
In the optically thick limit, the radiative flux in the comoving frame follows the diffusion approximation given by Equation~\eqref{eq:dif_app_E}.
Meanwhile, the total flux in the lab frame includes advection by the moving gas that cannot be neglected in the dynamic diffusion regime as shown in Equation~\eqref{eq_Lorentz:F2}.
Accordingly, when $\tau_\mr{cell}\gg1$ and the trapped component $E_\mathrm{rad,T}$ dominates, the radiative flux should be
\begin{eqnarray}
    && \bm{F}_\mr{T} = \bm{F}_\mr{T,diff} + \bm{F}_\mr{T,adv} \label{eq:F_T}, \\
    && \bm{F}_\mr{T,diff}= -\frac{c} {3\chi_\mathrm{0R}\rho}\nabla E_\mathrm{rad,T}, \label{eq:F_Tdiff} \\
    && \bm{F}_\mr{T,adv}  = \f{4\bm{v}}{3} E_\mr{rad,T}. \label{eq:F_Tadv}
\end{eqnarray}
Note that, based on Equation~\eqref{eq_app:EP_identical}, we use the lab frame quantity on the right-hand side of Equations~\eqref{eq:F_Tdiff} and \eqref{eq:F_Tadv}.
\par
When solving Equation~\eqref{eq_basic:Erad}, the flux term $\nabla\cdot\bm{F}$ is evaluated using the GLF Riemann solver, which includes the numerical flux given by Equation~\eqref{eq:Fnum}.  
Therefore, the diffusive part of the trapped component, $\bm{F}_{\mathrm{T,diff}}$, is automatically accounted for by the numerical diffusion of the solver, and we only need to explicitly add the advective contribution $\bm{F}_{\mathrm{T,adv}}$ to the streaming flux, leading to the modification
\begin{eqnarray}
    \bm{F}~\mathrm{in}~\eqref{eq_basic:Erad} \rightarrow \bm{F}_{\mathrm{S}} + \bm{F}_{\mathrm{T,adv}}.
\end{eqnarray}
In contrast, the source term $\bm{G}$ in Equation~\eqref{eq_basic:G} is, as is generally the case, directly evaluated without using a Riemann solver.
Therefore, to properly account for the contribution of both the diffusive and advective flux of the trapped component, we modify the flux in the source term as
\begin{eqnarray}
    \bm{F}~\mathrm{in}~\eqref{eq_basic:G} \rightarrow \bm{F}_{\mathrm{S}} + \bm{F}_{\mathrm{T,diff}}+ \bm{F}_{\mathrm{T,adv}}. \label{eq:Fmod_in_source}
\end{eqnarray}
Using Equations~\eqref{eq:F_Tdiff}, \eqref{eq:F_Tadv}, and \eqref{eq:Fmod_in_source}, together with the relation $E_\mathrm{rad}=E_\mathrm{rad,S}+E_\mathrm{rad,T}$, we can rewrite the source term $\bm{G}$ given by Equation~\eqref{eq_basic:G} as
\begin{eqnarray}
   && \rho\chi_{0F}\f{1}{c} \bra{\bm{F}_\mr{S}+\bm{F}_\mr{T,diff}+\bm{F}_\mr{T,adv}} -\rho\chi_{0F}\f{4\bm{v}}{3c}E_\mr{rad}, \nonumber \\
   && = \bra{\rho\chi_{0F}\f{\bm{F}_\mr{S}}{c}-\rho\chi_{0F}\f{4\bm{v}}{3c}E_\mr{rad,S}} -\f{\chi_{0F}}{3\chi_{0\mr{R}}} \nabla E_\mr{rad,T}, \nonumber \\ 
   && \simeq \bra{\rho\chi_{0F}\f{\bm{F}_\mr{S}}{c}-\rho\chi_{0F}\f{4\bm{v}}{3c}E_\mr{rad,S}} -\f{1}{3} \nabla E_\mr{rad,T}. \label{eq:Fmod_source}
\end{eqnarray}
We approximate $\chi_{0F}$ as $\chi_{0\mathrm{R}}$ when deriving the last expression, since $E_{\mathrm{rad,T}}$ dominates only in the optically thick regime.
With the above modifications, Equations~\eqref{eq_basic:Egas}--\eqref{eq_basic:G} can be rewritten as
\begin{eqnarray}
    && \f{\p E_\mr{gas}}{\p t} + \nabla\cdot\left\{\bm{v}(E_\mr{gas}+P_\mr{gas})\right\} = \rho\bm{g}\cdot\bm{v} + c G^0  ,\label{eq_ST:Egas} \\
    && \f{\p \rho\bm{v}}{\p t} + \nabla\cdot(\rho\bm{v}\otimes\bm{v}+P_\mr{gas}\ten{I}) = 
    \rho\bm{g} + \bm{G}_\mr{S} + \bm{G}_\mr{T} , \label{eq_ST:rhov} \\
    && \f{\p E_\mr{rad}}{\p t} + \nabla\cdot\bra{\bm{F}_\mr{S}+\f{4\bm{v}}{3}E_\mr{rad,T}} = -cG^0 , \label{eq_ST:Erad} \\
    && \f{\p \bm{F}_\mr{S}}{\p t} + c^2\nabla\cdot\ten{P}_\mr{rad,S} = - c^2 \bm{G}_\mr{S} , \label{eq_ST:F} \\
    && G^0 = \rho  ( \kappa_{0E} E_\mr{rad} - \kappa_\mr{0P} a T^4 ) , \label{eq_ST:G0} \\
    && \bm{G}_\mr{S} = \rho \chi_{0F} \f{\bm{F}_\mr{S}}{c} - \rho \chi_\mr{0F}\f{4\bm{v}}{3c} E_\mr{rad,S}  ,  \label{eq_ST:GS} \\
    && \bm{G}_\mr{T} = -\f{1}{3}\nabla E_\mathrm{rad,T} , \label{eq_ST:GT} 
\end{eqnarray}
where the vectors $\bm{G}_\mr{S}$ and $\bm{G}_\mr{T}$ represent the radiation force densities associated with the streaming and trapped photon components, respectively.
Note that Equation~\eqref{eq_ST:F} now describes the evolution of the streaming photon flux $\bm{F}_\mathrm{S}$, rather than the total radiative flux.
The radiation-pressure tensor $\ten{P}_\mr{rad,S}$ is calculated from $E_\mr{rad,S}$ and $\bm{F}_\mr{S}$ in the same manner as Equation~\eqref{eq:D},
\begin{eqnarray}
    && \ten{P}_\mr{rad,S} = \ten{D}_\mr{S}E_\mr{rad,S} , \nonumber \\
    && \ten{D}_\mr{S} = \f{1-\chirad_\mr{S}}{2}\ten{I} + \f{3\chirad_\mr{S}-1}{2}\bm{n}_\mr{S}\otimes\bm{n}_\mr{S}, \label{eq:D_S}\\
    && \bm{n}_\mr{S}=\f{\bm{F}_\mr{S}}{|\bm{F}_\mr{S}|}, \h
    \chirad_\mr{S}=\f{3+4(|\bm{F}_\mr{S}|/(cE_\mr{rad,S}))^2}{5+2\sqrt{4-3(|\bm{F}_\mr{S}|/(cE_\mr{rad,S}))^2}}
    . \nonumber 
\end{eqnarray}
These equations are equivalent to those presented in Appendix B of \citet{Rosdahl_and_Teyssier_2015}, except for some differences.
First, we have not yet applied the RSLA to these equations.
Next, we include scattering opacity, use lab frame quantities for the energy densities appearing in the radiation four-force terms, and retain only the dominant terms as described in Appendix~\ref{sec:app_dom}.
Furthermore, we do not evolve the streaming and trapped energy densities separately, but instead describe the time evolution of their sum using Equation~\eqref{eq_ST:Erad}.
This choice of evolving the combined quantity, rather than treating the two components individually, is merely a matter of formulation and does not represent a fundamental difference.
\subsection{Reduced speed of light approximation} \label{subsec:RSLA}
In RHD simulations, explicitly solving the advection terms $\nabla\cdot\bm{F}_\mr{S}$ and $c^2 \nabla\cdot\ten{P}_\mr{rad,S}$ in Equations~\eqref{eq_ST:Erad} and \eqref{eq_ST:F} requires a small time step of $\Delta t < \Delta x / 3c$ for numerical stability, which is the so-called CFL condition.
When the gas velocity is much smaller than the speed of light ($\beta \ll 1$), this condition leads to a prohibitively high computational cost for following the gas evolution.
The RSLA alleviates this issue by artificially reducing the speed of light by a factor $f_c$ as
\begin{eqnarray}
    \tilde{c} = f_c c \label{eq:ctilde} ,
\end{eqnarray} 
thereby relaxing the CFL condition to $\Delta t < \Delta x / 3\tilde{c}$.
\par
Practically, we implement the RSLA by multiplying the advection terms by the reduction factor $f_c$, thereby reducing their magnitude.
However, the specific formulation of the RSLA can vary depending on how $f_c$ is applied within the RHD equations \citep{Hopkins_et_al_2022}.
If the reduction is applied only to the advection terms, $\nabla\cdot\bm{F}$ and $c^2\nabla\cdot\ten{P}_\mr{rad,S}$, which are responsible for the restrictive CFL condition, the resulting steady-state solutions may differ from those of the original, nonreduced system.
To prevent this inconsistency, many previous studies have applied the reduction factor $f_c$ not only to the advection terms but also to the radiation source terms, $G^0$ and $\bm{G}$ \citep[e.g.,][]{Skinner_and_Ostriker_2013, Rosdahl_and_Teyssier_2015,Kannan_et_al_2019,Chang_et_al_2020,Fukushima_and_Yajima_2021}.
In this study, we adopt the same approach, except that we do not apply $f_c$ to the radiation advection term associated with gas dragging, $(4\bm{v}/3)E_\mr{rad,T}$ on the left-hand side of Equation~\eqref{eq_ST:Erad}.
This is because this term represents the advection together with the gas and therefore does not impose a restrictive time step.
Moreover, applying $f_c$ to this term would instead introduce an unphysical lag between radiation and gas motion, leading to unphysical behavior.
With this treatment, Equations~\eqref{eq_ST:Erad} and \eqref{eq_ST:F} are modified as follows:
\begin{eqnarray}
      && \f{\p E_\mr{rad}}{\p t} + \nabla\cdot\bra{ f_c \bm{F}_\mr{S}+\f{4\bm{v}}{3}E_\mr{rad,T}} = - f_c cG^0 , \label{eq_RSLA_pre:Erad}  \\
      && \f{\p \bm{F}_\mr{S}}{\p t} + f_c c^2\nabla\cdot\ten{P}_\mr{rad,S} = - f_c c^2 \bm{G}_\mr{S} .\label{eq_RSLA_pre:F} 
\end{eqnarray}
Under this RSLA formulation, it is convenient to define new radiation variables as
\begin{equation}
\begin{array}{l}
    {\displaystyle \EofRSLA_\mr{rad} = \f{E_\mr{rad}}{f_c}, 
    \hspace{3mm} \EofRSLA_\mr{rad,S} = \f{E_\mr{rad,S}}{f_c} , } \\ [10pt]
    {\displaystyle \EofRSLA_\mr{rad,T} = \f{E_\mr{rad,T}}{f_c},
    \hspace{3mm} \PofRSLA_\mr{rad} = \f{\ten{P}_\mr{rad}}{f_c} ,}
\end{array}
\label{eq:def_tildeErad}
\end{equation}
where, from Equation~\eqref{eq:distribute_ST}, $\EofRSLA_\mr{rad}$, $\EofRSLA_\mr{rad,S}$, and $\EofRSLA_\mr{rad,T}$ are related as
\begin{equation}
\begin{array}{l}
    \EofRSLA_\mr{rad,S} = (1-\fdis_\mr{T}) \EofRSLA_\mr{rad} , \hspace{3mm} \EofRSLA_\mr{rad,T} = \fdis_\mr{T} \EofRSLA_\mr{rad} , \\ [6pt]
    {\displaystyle \fdis_\mr{T} = \exp\left(-\frac{2}{3\tau_\mr{cell}}\right) .}
\end{array}
\label{eq:distributeRSLA_ST}
\end{equation}
Here, $\EofRSLA_\mr{rad}$ and $\PofRSLA_\mr{rad}$ are identical to $E$ and $\bm{P}$ given by Equations~(2) and (4) in \citet{Rosdahl_and_Teyssier_2015}.
With these variables, Equations~\eqref{eq_ST:Egas}, \eqref{eq_ST:rhov}, \eqref{eq_ST:G0}--\eqref{eq_ST:GT}, \eqref{eq_RSLA_pre:Erad} and \eqref{eq_RSLA_pre:F} can be rewritten as 
\begin{eqnarray}
    && \f{\p E_\mr{gas}}{\p t} + \nabla\cdot\left\{\bm{v}(E_\mr{gas}+P_\mr{gas})\right\} = \rho\bm{g}\cdot\bm{v} + c G^0  ,\label{eq_RSLA:Egas} \\
    && \f{\p \rho\bm{v}}{\p t} + \nabla\cdot(\rho\bm{v}\otimes\bm{v}+P_\mr{gas}\ten{I}) = 
    \rho\bm{g} + \bm{G}_\mr{S} + \bm{G}_\mr{T} , \label{eq_RSLA:rhov} \\
    && \f{\p \EofRSLA_\mr{rad}}{\p t} + \nabla\cdot\bra{\bm{F}_\mr{S}+\f{4\bm{v}}{3}\EofRSLA_\mr{rad,T}} = -cG^0 , \label{eq_RSLA:Erad} \\
    && \f{\p \bm{F}_\mr{S}}{\p t} + \tilde{c}^2 \nabla\cdot\PofRSLA_\mr{rad,S} = - c \tilde{c} \bm{G}_\mr{S} , \label{eq_RSLA:F} \\
    && G^0 = \rho  ( \kappa_{0E} f_c \EofRSLA_\mr{rad} - \kappa_\mr{0P} a T^4 ) , \label{eq_RSLA:G0} \\
    && \bm{G}_\mr{S} = \rho \chi_{0F} \f{\bm{F}_\mr{S}}{c} - \rho \chi_\mr{0F}\f{4\bm{v}}{3c} f_c \EofRSLA_\mr{rad,S}  ,  \label{eq_RSLA:GS} \\
    && \bm{G}_\mr{T} = -\f{1}{3}\nabla f_c \EofRSLA_\mathrm{rad,T} . \label{eq_RSLA:GT} 
\end{eqnarray}
From Equations~\eqref{eq_RSLA:Egas} and \eqref{eq_RSLA:Erad}, it is evident that the sum $E_\mathrm{gas} + \EofRSLA_\mathrm{rad}$ is conserved during the gas-radiation interaction under the RSLA.
The radiation-pressure tensor $\PofRSLA_\mr{rad,S}$ is calculated from $\EofRSLA_\mr{rad,S}$ and $\bm{F}_\mr{S}$ as
\begin{eqnarray}
    && \PofRSLA_\mr{rad,S} = \ten{D}_\mr{S}\EofRSLA_\mr{rad,S} , \nonumber \\
    && \ten{D}_\mr{S} = \f{1-\chirad_\mr{S}}{2}\ten{I} + \f{3\chirad_\mr{S}-1}{2}\bm{n}_\mr{S}\otimes\bm{n}_\mr{S}, \label{eq:tilde_D_S}\\
    && \bm{n}_\mr{S}=\f{\bm{F}_\mr{S}}{|\bm{F}_\mr{S}|}, \h
    \chirad_\mr{S}=\f{3+4(|\bm{F}_\mr{S}|/(\tilde{c}\EofRSLA_\mr{rad,S}))^2}{5+2\sqrt{4-3(|\bm{F}_\mr{S}|/(\tilde{c}\EofRSLA_\mr{rad,S}))^2}}
    , \nonumber 
\end{eqnarray}
which is in the same form as Equations~\eqref{eq:D} and \eqref{eq:D_S}. 
\section{Challenges of RHD simulations that resolve the protostellar interior} \label{sec:challenges}
In this section, we discuss the three key challenges encountered when applying the M1 closure method with the RSLA to star formation RHD simulations that resolve the protostellar interior.
We present solutions to each of these challenges in Section~\ref{sec:improved_scheme}.
\subsection{Challenge 1: energy conservation with the RSLA} \label{subsec:energy-nonconserve}
While the formulation of the RSLA introduced in Section~\ref{subsec:RSLA} is widely adopted to improve computational efficiency, it induces the nonconservation of total physical energy.
Actually, as can be seen from Equations~\eqref{eq_RSLA_pre:Erad} and \eqref{eq_RSLA:Egas}, when $G^0 < 0$, the gas loses an amount of energy $-cG^0$ through its interaction with radiation, whereas the radiation gains only $-f_c cG^0$.
This implies that, whenever the energy is transferred from the gas to the radiation with $G^0<0$, the radiation field receives only a fraction $f_c$ of the energy it would physically acquire, and the remaining fraction $(1-f_c)$ is artificially removed from the total energy budget.
\par
In optically thin systems, this nonconservation is generally not problematic, as radiation energy is lost from the system regardless of whether the RSLA is applied or not.
Moreover, since both the rate of energy exchange $cG^0$ and the transport $\nabla\cdot\bm{F}$ are reduced by the same factor $f_c$ as shown in Equation~\eqref{eq_RSLA_pre:Erad}, the resulting steady-state radiation flux $\bm{F}$ remains consistent with that obtained without applying the RSLA. Consequently, radiation feedback effects such as radiation pressure can still be accurately modeled in this regime.
\par
In contrast, in optically thick systems, when the temperature is sufficiently high, the radiation energy becomes dominant within the medium.
When this energy is artificially lost due to the RSLA, we cannot accurately follow the thermal and dynamical evolution of the system \citep[see, e.g., the test problem in][]{Wibking_et_al_2022}.
In Section~\ref{subsec:HybridRT}, we introduce a new scheme designed to restore physical energy conservation in the optically thick limit.
\subsection{Challenge 2: radiative transfer with steep optical-depth gradient} \label{subsec:RT_in_transition}
As mentioned in Section~\ref{subsec:RT15}, the original M1 closure scheme fails to solve RT in regions where the optical depth per cell exceeds unity due to excessive numerical diffusion. 
Then, to accurately follow the RT in such a regime, we introduce the method by \citet{Rosdahl_and_Teyssier_2015}, which splits the radiation energy $\EofRSLA_\mr{rad}$ into streaming component $\EofRSLA_\mr{rad,S}$ and trapped component $\EofRSLA_\mr{rad,T}$.
However, even with this method, a problem in RT still remains.
\par
As an example, consider a situation in which an optically thick cell with $\tau_\mathrm{cell} \gg 1$ is directly adjacent to an optically thin cell with $\tau_\mathrm{cell} \ll 1$.
Physically, radiation energy should escape from the surfaces of the optically thick cell at a flux of $\sigma_\mathrm{SB} T^4$, 
where $\sigma_\mr{SB}$ is the Stefan–Boltzmann constant.
However, when we use the method introduced in Section~\ref{subsec:RT15}, the trapped component $\EofRSLA_\mr{rad,T}$ dominates, and the streaming component $\EofRSLA_\mr{rad,S}$ is suppressed to very small values in optically thick cells.
Since we calculate RT only with $\EofRSLA_\mathrm{rad,S}$, radiation cannot effectively escape from the optically thick cell.
\par
This situation is particularly relevant at the surface of a protostar, where the density and opacity vary steeply over short spatial scales.
For a solar-mass main-sequence star, this scale is only on the order of $10^2$~km.
These scales are far too small to be resolved in simulations that follow the global evolution of the entire star.
However, accurately solving RT across such boundaries is essential, because it determines how efficiently the protostar loses entropy through radiation and how much radiative feedback is imparted to the surrounding gas.
We address this issue in Section~\ref{subsec:RadPartition}.
\subsection{Challenge 3: modeling the radiation spectrum}\label{subsec:overview_spectrum}
Since the gas opacity to radiation is highly sensitive to frequency, the interaction between gas and radiation, such as photoheating, photoionization, and radiation pressure, depends strongly on the shape of the radiation spectrum. 
In this study, we also estimate the mean opacity with the spectrum as in Equations~\eqref{eq:kappa_0E} and \eqref{eq:kappa_0F}.
While the most straightforward method for accurately following the evolution of the spectrum is solving the RHD equations for many frequency bins, their computational cost scales linearly with the number of bins and becomes prohibitively expensive.
Instead, most RHD simulations employ only one (gray approximation) or a few frequency bins, and assume a spectral shape within each bin \citep{Skinner_and_Ostriker_2013,Mignon-Risse_et_al_2020,Wibking_et_al_2022}.
\par
Among various approaches to modeling the spectral shape, one of the simplest is to approximate the spectrum in each cell by a Planck function at the local gas temperature.
Although this approach is valid when the system is entirely optically thick, it fails to capture the spectral evolution in optically thin regions and leads to inaccurate estimates of radiative feedback in such environments.
\par
Alternatively, in simulations that employ sink particles to represent individual stars or stellar clusters, the spectral shape is often determined based on 1D stellar evolution models. 
However, such an approach is not applicable to simulations like ours, in which stars are resolved directly during the course of the calculation, and the radiation spectrum must be determined consistently with their 3D evolution.
Moreover, the above method cannot capture spatial variations in the radiation spectrum.
This challenge is not specific to the M1 scheme but arises with any closure relation, including the variable Eddington tensor method, and is therefore a general difficulty of RHD simulations.
In Section~\ref{sec:spectral_reconstruction}, we describe our method for capturing the time evolution and spatial variation of the spectral shape.
\section{Improved Scheme} \label{sec:improved_scheme}
In this section, we present our solutions to the challenges outlined in Section~\ref{sec:challenges}.
Section~\ref{subsec:HybridRT} describes how we address the violation of energy conservation caused by the RSLA.
In Section~\ref{subsec:RadPartition}, we introduce a method for accurately solving RT in systems with a steep optical-depth gradient.
Finally, in Section~\ref{sec:spectral_reconstruction}, we describe our approach for evaluating the radiation spectrum.
For the convenience of the readers, before proceeding to the description of the improved scheme, we first summarize the notation and relations for the constants, gas variables, and radiation variables used in the improved scheme in Tables~\ref{tab:summary_notation_constant}, \ref{tab:summary_notation_gas}, and \ref{tab:summary_notation_radiation}.
\begin{table}
  \centering
  \caption{Notation for constants}
  \label{tab:summary_notation_constant}
  \begin{tabular}{ll}
    \hline
    Symbol & Meaning \\
    \hline
    $k_\mr{B}$ & Boltzmann constant \\
    $m_\chH$ & mass of a hydrogen nucleus \\
    $h$ & Planck constant \\
    $\sigma_\mr{SB}$ & Stefan–Boltzmann constant \\
    $a$ & radiation constant \\
    $G$ & gravitational constant \\
    $c$ & speed of light \\
    $f_c$ & reduction factor in the RSLA \\
    $\tilde{c}=f_c c$ & reduced speed of light \\
    $\Delta x$ & computational cell width \\
    $\Delta t$ & computational time step \\
    \hline
  \end{tabular}
\end{table}
\begin{table}
  \centering
  \caption{Notation for gas variables}
  \label{tab:summary_notation_gas}
  \begin{tabular}{ll}
    \hline
    Symbol & Meaning \\
    \hline
    $\rho$ & gas density \\
    $P_\mr{gas}$ & gas pressure \\
    T & gas temperature \\
    $\bm{v}$ & gas velocity \\
    $\beta = |\bm{v}|/c$ & ratio of gas speed to light speed \\
    $c_\mr{s}$ & sound speed \\
    $\mu$ & mean molecular weight \\
    $\gamma$ & specific heat ratio \\
    $E_\mr{gas}$ & gas total energy \\
    $\phi$ & gravitational potential \\
    $\bm{g}$ & gravitational acceleration \\
    \hline
  \end{tabular}
\end{table}
%
%
%
\begin{table*}
  \centering
  \caption{Notation and relations for radiation variables. For simplicity, variables related to the photon-number density are omitted.}
  \label{tab:summary_notation_radiation}
  \begin{tabular}{lll}
    \hline\hline
    Symbol & Relation & Meaning \\
    \hline\hline

    \multicolumn{3}{c}{Energy-density variables} \\
    \hline
    $E^\mr{con}_\mr{rad}$ 
      & $E^\mr{con}_\mr{rad} = E^\mr{R}_\mr{rad} + E^\mr{N}_\mr{rad}$ 
      & total conserved radiation energy-density variable \\

    $E^\mr{R}_\mr{rad}$ 
      & $E^\mr{R}_\mr{rad} = (1-\eta^\mr{N}) E^\mr{con}_\mr{rad} =  E^\mr{R,S}_\mr{rad} + E^\mr{R,T}_\mr{rad}$ 
      & reduced speed of light approximation (RSLA) component \\
      &
      & (scaled by $1/f_c$ from the physical energy density) \\

    $E^\mr{N}_\mr{rad}$ 
      & $E^\mr{N}_\mr{rad} = \eta^\mr{N} E^\mr{con}_\mr{rad}$ 
      & non-RSLA (full speed of light) component \\
      &
      & (corresponding to the physical energy density) \\

    $E^\mr{R,S}_\mr{rad}$ 
      & $E^\mr{R,S}_\mr{rad} = (1-\eta_\mr{T}) E^\mr{R}_\mr{rad}$ 
      & streaming component of the RSLA radiation field \\

    $E^\mr{R,T}_\mr{rad}$ 
      & $E^\mr{R,T}_\mr{rad} = \eta_\mr{T} E^\mr{R}_\mr{rad}$ 
      & trapped component of the RSLA radiation field \\

    \hline
    \multicolumn{3}{c}{Fraction and optical depth per computational cell} \\
    \hline

    $\eta^\mr{N}$ 
      & $\eta^\mr{N} = \exp\!\left(-\frac{2}{3 f_c \tau_\mr{cell}}\right)$ 
      & non-RSLA fraction of total radiation field \\

    $\eta_\mr{T}$ 
      & $\eta_\mr{T} = \exp\!\left(-\frac{2}{3 \tau_\mr{cell}}\right)$ 
      & trapped fraction of the RSLA radiation field \\

    $\tau_\mr{cell}$ 
      & $\tau_\mr{cell} = \rho \chi_\mr{0R} \Delta x$ 
      & optical depth per computational cell \\

       \hline
    \multicolumn{3}{c}{RSLA-streaming transport and closure variables} \\
    \hline

    $\bm{F}_\mr{S}$ 
      &  
      & radiation flux of the RSLA-streaming component \\

    $\PofRSLA_\mr{rad,S}$ 
      & $\PofRSLA_\mr{rad,S} = \ten{D}_\mr{S} \EofRSLA_\mr{rad,S}$ 
      & radiation-pressure tensor of the RSLA-streaming component \\

    $\ten{D}_\mr{S}$ 
      & 
      & Eddington tensor for the RSLA-streaming component \\

    \hline
    \multicolumn{3}{c}{Radiation four-force density and decomposition} \\
    \hline

    $G^0$ 
      & 
      & time component of radiation four-force density \\

    $\bm{G}_\mr{tot}$ 
      & $\bm{G}_\mr{tot} = \bmGofRSLA_\mr{S} + \bmGofRSLA_\mr{T} + \bmGofNonRSLA$ 
      & total spatial component of radiation four-force density \\

    $\bmGofRSLA_\mr{S}$ 
      &  
      & RSLA-streaming component of $\bm{G}_\mr{tot}$ \\

    $\bmGofRSLA_\mr{T}$ 
      & 
      & RSLA-trapped component of $\bm{G}_\mr{tot}$  \\

    $\bmGofNonRSLA$ 
      & 
      & non-RSLA component of $\bm{G}_\mr{tot}$ \\

    \hline
    \multicolumn{3}{c}{Opacity and spectral variables} \\
    \hline

    $\kappa^\nu_0$ 
      & 
      & frequency-dependent absorption opacity in the comoving frame \\

    $\sigma^\nu_0$ 
      & 
      & frequency-dependent scattering opacity in the comoving frame \\

    $\kappa_\mr{0P}$ 
      & 
      & Planck mean absorption opacity \\

    $\kappa_{0E}$ 
      & 
      & energy-mean absorption opacity \\

    $\chi_{0E}$ 
      & 
      & energy-mean total (absorption + scattering) opacity \\

    $\chi_{0F}$ 
      & 
      & flux-mean total opacity \\

    $\chi_\mr{0R}$ 
      & 
      & Rosseland mean opacity \\

    $\chi_\mr{crit}$ 
      & 
      & critical opacity above which the non-RSLA component dominates \\

    $\hat{\chi}$ 
      & $\hat{\chi} = \max(\chi_{0R}, \chi_\mr{crit})$ 
      & effective opacity for diffusive transport of the non-RSLA component \\

    $B^\nu(T)$ 
      & 
      & Planck function at temperature $T$ \\

    $E^\nu_{\mr{rad},0}$ 
      & 
      & comoving-frame radiation energy density per unit frequency \\

    $F^\nu_{0}$ 
      & 
      & comoving-frame radiation flux per unit frequency \\

    $T_\mr{rad}$ 
      & 
      & radiation temperature defined from the mean photon energy \\

    \hline\hline
  \end{tabular}
\end{table*}
\subsection{Hybrid radiative transfer: RSLA and Non-RSLA components} \label{subsec:HybridRT}
As discussed in Section~\ref{subsec:energy-nonconserve}, applying the RSLA violates physical energy conservation, making it difficult to accurately follow the time evolution of optically thick systems, especially when the radiation energy dominates over the gas internal energy.
However, in very optically thick regions, radiation propagates slowly via diffusion, and therefore, it is possible to evolve the radiation without applying the RSLA while maintaining a reasonably large time step.
Indeed, considering the diffusion approximation given in Equation~\eqref{eq:dif_app_E}, von Neumann stability analysis requires the time step to satisfy
\begin{eqnarray}
\Delta t < \frac{\chi_\mr{0R} \rho \Delta x^2}{2c} .
\end{eqnarray}
In regions where the opacity satisfies $\chi_\mr{0R} > \chi_\mr{crit}\equiv2 / (3f_c \rho \Delta x)$, this condition is less restrictive than the CFL condition for the M1 closure scheme with the RSLA, which is $\Delta t < \Delta x / 3\tilde{c}$.
This fact indicates that we can evolve radiation without applying the RSLA in such regions.
\par
Motivated by this, we introduce an additional radiation energy component, denoted as $\EofNonRSLA_\mathrm{rad}$, which is evolved without the RSLA and dominates in regions where $\chi_\mr{0R}\geq \chi_\mr{crit}$. 
Before presenting the detailed formulation, we first summarize the three radiation components used throughout this paper.
First, we consider two components to which the RSLA is applied:
\\ \par
(1) \textit{RSLA–streaming} component, $\EofRSLA_{\mathrm{rad,S}}$. This component represents the portion of the radiation field subject to the RSLA that is transported using the standard M1 closure.
\\ \par
(2) \textit{RSLA–trapped} component, $\EofRSLA_{\mathrm{rad,T}}$. This component corresponds to the remaining part of the RSLA radiation field that is not transported by the M1 scheme.
\\
\\
The total RSLA radiation energy is given by $\EofRSLA_\mathrm{rad}=\EofRSLA_\mathrm{rad,S}+\EofRSLA_\mathrm{rad,T}$.
As noted in Section~\ref{subsec:RSLA}, the conserved quantity of the RSLA component in the gas–radiation interaction is the radiation energy density divided by the reduction factor $f_c$, as defined in Equation~\eqref{eq:def_tildeErad}. 
The $\EofRSLA_\mathrm{rad}$ corresponds to this conserved quantity.
Next, we consider the third component, to which the RSLA is not applied:
\\ \par
(3) \textit{Non-RSLA} component, $\EofNonRSLA_{\mathrm{rad}}$. This component is transported diffusively with the true speed of light.
\\
\\
Because the non-RSLA component evolves at the full speed of light, its radiation energy $\EofNonRSLA_\mathrm{rad}$ is the conserved quantity.
Therefore, in the region where $\chi_\mr{0R}\geq\chi_\mr{crit}$ and $\EofNonRSLA_\mathrm{rad}$ dominates, we can accurately follow the evolution of optically thick systems while conserving energy.
\par
To ensure that $\EofNonRSLA_\mathrm{rad}$ dominates only in regions where $\chi_\mr{0R} \geq \chi_\mr{crit}$, we define the total conserved radiation energy as 
\begin{eqnarray}
    E^\mathrm{con}_\mathrm{rad} = \EofNonRSLA_\mathrm{rad} + \EofRSLA_\mathrm{rad} , \label{eq:Econ}
\end{eqnarray}
and redistribute it between the two components during the simulation such that
\begin{eqnarray}
\begin{array}{l}
    \EofRSLA_\mr{rad}= (1-\fdisofNonRSLA)E^\mr{con}_\mr{rad}, \hspace{3mm} \EofNonRSLA_\mr{rad}=\fdisofNonRSLA E^\mr{con}_\mr{rad}, \\ [6pt]
    {\displaystyle \fdisofNonRSLA = \exp\left(-\f{\chi_\mr{crit}}{\chi_\mr{0R}}\right) =\exp\left(-\f{2}{3f_c \tau_\mr{cell}}\right) ,} 
\end{array}
\label{eq:distribute_RN}
\end{eqnarray}
during the calculation. The timing of this redistribution can be chosen appropriately depending on the specific implementation of the code (see Section~\ref{sec:comp_proc}).
With this formulation, if an optically thick system contains regions where $\tau_\mathrm{cell} < 2 / (3f_c)$, local energy conservation breaks down in such regions, since the RSLA component dominates. 
We discuss this impact in Section~\ref{subsec:Local_break}.
\par
Since the non-RSLA component $\EofNonRSLA_\mr{rad}$ dominates in optically thick regions where $\chi_\mr{0R}\geq\chi_\mr{crit}$, we model its transport as
\begin{eqnarray}
    && \bmFofNonRSLA=\bmFofNonRSLA_\mr{diff} + \bmFofNonRSLA_\mr{adv} , \label{eq:Fhat} \\  
    && \bmFofNonRSLA_\mr{diff} = -\f{c}{3\hat{\chi}\rho} \nabla \EofNonRSLA_\mr{rad}, \hspace{3mm} \hat{\chi} = \max(\chi_\mr{0R},\chi_\mr{crit}), \label{eq:Fhat_diff} \\
    && \bmFofNonRSLA_\mr{adv} = \f{4\bm{v}}{3}\EofNonRSLA_\mr{rad}.
\end{eqnarray}
In Equation~\eqref{eq:Fhat}, $\bmFofNonRSLA_\mr{diff}$ represents radiative diffusion, while $\bmFofNonRSLA_\mr{adv}$ corresponds to advection by the gas flow (see Equation~\ref{eq_Lorentz:F2}).
As mentioned above, the von Neumann stability analysis ensures that the radiation diffusion can be stably solved using a time step that satisfies the CFL condition imposed by the M1 closure scheme with the RSLA.
\par
We now present the RHD equations incorporating the non-RSLA radiation component $\EofNonRSLA_\mathrm{rad}$.
These equations are derived by extending Equations~\eqref{eq_RSLA:Egas}--\eqref{eq_RSLA:GT} to include the non-RSLA component whose flux is given by Equation~\eqref{eq:Fhat}.
The resulting equations can be written as
\begin{eqnarray}
    && \f{\p E_\mr{gas}}{\p t} + \nabla\cdot\left\{\bm{v}(E_\mr{gas}+P_\mr{gas})\right\} = \rho\bm{g}\cdot\bm{v}
    + cG^0 , \label{eq:E_gas} \\
    && \f{\p \rho\bm{v}}{\p t} + \nabla\cdot(\rho\bm{v}\otimes\bm{v}+P_\mr{gas}\bm{I}) = \rho\bm{g} +\bmGofRSLA_\mr{S} + \bmGofRSLA_\mr{T} + \bmGofNonRSLA , \nonumber \\ \label{eq:rhov} \\
    && \f{\p E^\mr{con}_\mr{rad}}{\p t} + \nabla\cdot \bigg\{ \bm{F}_\mr{S} - \f{c}{3\hat{\chi}\rho}\nabla \EofNonRSLA_\mr{rad} + \f{4\bm{v}}{3}(\EofRSLA_\mr{rad,T}+\EofNonRSLA_\mr{rad}) \bigg\} \nonumber \\ 
    && \hspace{6cm} = -cG^0 , \label{eq:tildeE} \\
    && \f{\p \bm{F}_\mr{S}}{\p t} + \tilde{c}^2\nabla\cdot\PofRSLA_\mr{rad,S} = -c\tilde{c} \bmGofRSLA_\mr{S} , \label{eq:tildeF} \\
    && G^0 = \rho \left\{  \kappa_{0E} (f_c \EofRSLA_\mr{rad} + \EofNonRSLA_\mr{rad}) - \kappa_\mr{0P} a T^4 \right\} , \label{eq:tildeG0} \\
    && \bmGofRSLA_\mr{S} = \rho \chi_{0F} \f{\bm{F}_\mr{S}}{c} - \rho \chi_\mr{0F}\f{4\bm{v}}{3c} f_c \EofRSLA_\mr{rad,S} \label{eq:tildeG} , \\
    && \bmGofRSLA_\mr{T} = -  \f{1}{3} \nabla f_c \EofRSLA_\mathrm{rad,T} ,
    \hspace{5mm} \bmGofNonRSLA = -  \f{1}{3} \nabla \EofNonRSLA_\mathrm{rad} , \label{eq:hatG} 
\end{eqnarray}
where $\bmGofRSLA_\mr{S}$ is the radiation force associated with the RSLA-streaming component, $\bmGofRSLA_\mr{T}$ the radiation-pressure gradient arising from the RSLA-trapped component, and $\bmGofNonRSLA$ the radiation-pressure gradient from the non-RSLA diffusive component.
The total spatial component of radiation four-force density is given by the sum of these contributions, $\bm{G}_\mr{tot}=\bmGofRSLA_\mr{S}+\bmGofRSLA_\mr{T}+\bmGofNonRSLA$.
Note that we do not evolve the three radiation components, $\EofRSLA_\mr{rad,S}$, $\EofRSLA_\mr{rad,T}$, and $\EofNonRSLA_\mr{rad}$, individually, but rather their sum, $E_{\rm rad}^{\rm con}$ with Equation~\eqref{eq:tildeE}.
From Equations~\eqref{eq:E_gas} and \eqref{eq:tildeE}, we can see that the $E_\mathrm{gas} + E^\mr{con}_\mr{rad}$ is conserved during the gas-radiation interaction.
We solve Equations~\eqref{eq:E_gas}--\eqref{eq:hatG} in combination with Equations~\eqref{eq_basic:rho}, \eqref{eq:distributeRSLA_ST}, \eqref{eq:tilde_D_S}, \eqref{eq:Econ}, and \eqref{eq:distribute_RN}.
\subsection{Radiation distribution based on neighboring cell information} \label{subsec:RadPartition}
We distribute the conserved quantity $E^\mr{con}_\mr{rad}$ among $\EofRSLA_\mathrm{rad,S}$, $\EofRSLA_\mathrm{rad,T}$, and $\EofNonRSLA_\mathrm{rad}$ based on the optical depth per cell $\tau_\mr{cell}$, as described in Equations~\eqref{eq:distribute_ST} and \eqref{eq:distribute_RN}.
However, as mentioned in Section~\ref{subsec:RT_in_transition}, this approach fails to capture the leakage of radiation from optically thick cells into neighboring optically thin cells, because the radiation energy in an optically thick cell is stored in either $\EofRSLA_\mathrm{rad,T}$ or $\EofNonRSLA_\mathrm{rad}$, neither of which is allowed to freely leave the cell. 
As a result, we cannot capture the outgoing radiation flux from optically thick cell surfaces.
This issue is particularly relevant to the narrow layers near the stellar surface, where the gas density and opacity rise steeply.
\par
To mitigate this problem, we use the minimum values of the optical depth per cell among the local and neighboring cells when computing $\fdis_\mathrm{T}$ and $\fdisofNonRSLA$.
By adopting this approach, when a neighboring cell is optically thin, radiation energy can be assigned to the streaming RSLA component $\EofRSLA_\mathrm{rad,S}$ even within an optically thick cell.
In such cases, when local emission and absorption are balanced, the physical radiation energy $f_c \EofRSLA_\mathrm{rad,S}$ approximately reaches the equilibrium value of $\sim a T^4$.
This energy can then propagate outward at the reduced speed of light $\tilde{c}$, resulting in a radiative flux per unit area per unit time estimated as
\begin{eqnarray}
\tilde{c} \EofRSLA_\mathrm{rad,S} \sim \tilde{c} \bra{\f{a T^4}{f_c}} \sim \sigma_\mathrm{SB} T^4,
\end{eqnarray}
which approximately reproduces the physically expected surface flux.
We describe in Section~\ref{sec:comp_proc} the details of when and how the optical depths of neighboring cells are incorporated into the calculation, together with the code implementation.
\subsection{Spectral reconstruction and opacity estimation from mean photon energy} \label{sec:spectral_reconstruction}
As mentioned in Section~\ref{subsec:overview_spectrum}, to accurately capture the interaction between the gas and radiation, we need to know the radiation spectrum.
However, the radiation energy density alone does not uniquely determine the spectral shape.
To address this issue, we evolve not only the radiation energy densities, $\EofRSLA_\mathrm{rad}$ and $\EofNonRSLA_\mathrm{rad}$, but also the corresponding photon-number densities, $\NofRSLA_\mathrm{rad}$ and $\NofNonRSLA_\mathrm{rad}$.
A similar approach has been adopted in several studies in numerical relativity, where both energy and number densities are evolved to reconstruct the neutrino spectrum \citep{Foucart_et_al_2016, Radice_et_al_2022, Musolino_et_al_2024}.
However, this method has not yet been applied to RT.
Analogous to the equations for the radiation energy density, we assume that the photon-number density evolves according to equations given in Appendix~\ref{sec:NumberDensity}.
\par
We reconstruct the local spectrum and estimate the opacities as follows.
First, we define the mean photon energy in each cell as
\begin{eqnarray}
    \langle h \nu \rangle = \frac{f_c \EofRSLA_\mathrm{rad} + \EofNonRSLA_\mathrm{rad}}{f_c \NofRSLA_\mathrm{rad} + \NofNonRSLA_\mathrm{rad}} , \label{eq:hnu_ave}
\end{eqnarray}
where $h$ is the Planck constant.
Note that we use the \textit{physical} radiation energy $f_c \EofRSLA_\mr{rad} + \EofNonRSLA_\mr{rad}$, rather than the conserved quantity $\EofRSLA_\mr{rad}+\EofNonRSLA_\mr{rad}$. 
We then introduce an effective radiation temperature $T_\mathrm{rad}$, which is defined consistently with the mean photon energy as
\begin{eqnarray}
    \langle h \nu \rangle = \frac{ \int_0^\infty d\nu  \, B_\nu(T_\mathrm{rad})}{ \int_0^\infty d\nu \, B_\nu(T_\mathrm{rad})/ h \nu } = \f{\pi^4}{30\zeta(3)}k_\mr{B}T_\mr{rad} ,  \label{eq:get_Trad}
\end{eqnarray}
where $\zeta(3)\simeq 1.202$ is the Riemann zeta function evaluated at 3.
Assuming that $E^\nu_\mathrm{rad,0}$ is proportional to the Planck distribution with a radiation temperature $T_\mathrm{rad}$, we can then evaluate the mean opacities, $\kappa_{0E}$ and $\kappa_{0E}^\prime$, which represent the opacities averaged over the spectra of radiation energy density and photon-number density, respectively, using Equations~\eqref{eq:kappa_0E} and \eqref{eq:kappa_0E_N}.
\par
Meanwhile, to evaluate the flux-mean opacity $\chi_{0F}$ and  $\chi_{0F}^\prime$ given by Equations~\eqref{eq:kappa_0F} and \eqref{eq:kappa_0F_N}, we need the flux spectrum $F^\nu_{0}$, which in general has a form different from that of $E^\nu_\mr{rad,0}$.
Then, we examine its behavior in the optically thin and thick limits.
In the optically thin limit, where the radiation flux $F^\nu_{0}$ approaches $c E^\nu_{\mr{rad},0}$, $\chi_{0F}$ can be approximated by
\begin{eqnarray}
    \chi_\mr{0E} = \f{\int_0^\infty d\nu \h (\kappa^\nu_{0} + \sigma^\nu_{0}) E^\nu_{\mr{rad},0}}{\int_0^\infty d\nu \h E^\nu_{\mr{rad},0}} . \label{eq:chi0E}
\end{eqnarray}
In the optically thick limit, $F^\nu_{0}$ follows the diffusion approximation, and $\chi_{0F}$ reduces to the Rosseland mean opacity given by Equation~\eqref{eq:Rosseland}.
To smoothly interpolate between these two limits, we adopt the following expression:
\begin{eqnarray}
\begin{array}{l}
    {\displaystyle \chi_{0F} = \f{3(1 - \xi_\chi)}{2} \chi_\mr{0R} + \f{3\xi_\chi - 1}{2} \chi_{0E} , } \\ [10pt]
    {\displaystyle \xi_\chi=\f{3+4[|\bm{F}_S|/\{c(f_c\EofRSLA_\mr{rad}+\EofNonRSLA_\mr{rad})\}]^2}{5+2\sqrt{4-3[|\bm{F}_S|/\{c(f_c\EofRSLA_\mr{rad}+\EofNonRSLA_\mr{rad})\}]^2}} ,}
\end{array}
\label{eq:xi_chi}
\end{eqnarray}
which is motivated by Equation~\eqref{eq:D}.
We estimate the opacity $\chi^\prime_{0F}$ with respect to the number density in the same manner, using Equation~\eqref{eq:kappa_0F_N_exp} in Appendix~\ref{sec:NumberDensity}.
This expression allows us to compute $\chi_{0F}$ and $\chi^\prime_{0F}$ based on the radiation spectrum $E^0_{\mr{rad},\nu}$ and the gas temperature $T$.
\par
For clarity, we summarize the opacities used in the RHD equations.
Energy exchange between gas and radiation (Equation~\ref{eq:tildeG0}) uses the Planck mean opacity $\kappa_\mr{0P}$ and the energy-mean opacity $\kappa_{0E}$ given by Equations~\eqref{eq:kappa_0P} and \eqref{eq:kappa_0E}.
Momentum exchange between the gas and the RSLA-streaming component (Equation~\ref{eq:tildeG}) uses the flux-mean opacity $\chi_{0F}$, whose behavior transitions smoothly according to the optical depth following Equation~\eqref{eq:xi_chi}.
In the optically thin limit, $\chi_{0F}$ reduces to the energy-mean opacity $\chi_{0E}$ (Equation~\ref{eq:chi0E}), whereas in the optically thick diffusion limit it approaches the Rosseland mean opacity $\chi_{0\mathrm{R}}$ (Equation~\ref{eq:Rosseland}), consistent with the diffusion approximation.
The quantity $\hat{\chi}$ appearing on the left-hand side of Equation~\eqref{eq:tildeE} governs the transport of the non-RSLA component in optically thick regions, and is therefore evaluated using the Rosseland mean opacity, as defined in Equation~\eqref{eq:Fhat_diff}.
The corresponding opacities for photon-number evolution are computed in an analogous manner.
\par
Finally, we note that the present spectral reconstruction implicitly assumes that the radiation field in each cell can be represented by a single-temperature, Planck-like spectrum. 
The method accurately reproduces spectral shapes, including blackbody emission from the protostar and thermal emission from dust. 
By self-consistently tracking the mean photon energy in space and time, the scheme enables radiative feedback processes, such as heating and radiation pressure, to be evaluated with good accuracy.

\section{Computational Procedure} \label{sec:comp_proc}
We implement the RHD scheme described in Section~\ref{sec:improved_scheme} into the self-gravitating magnetohydrodynamics code with adaptive mesh refinement, SFUMATO \citep{Matsumoto_2007}.
Several updated versions of SFUMATO have been developed to incorporate RT, including SFUMATO-RT \citep{Sugimura_et_al_2020}, which is based on ray tracing, and SFUMATO-M1 \citep{Fukushima_and_Yajima_2021}, which employs the M1 closure method combined with the RSLA.
These extensions have been applied to a variety of astrophysical simulations, and the implementation presented here further broadens the code's applicability.
\par
We solve Equations~\eqref{eq_basic:rho}, \eqref{eq:E_gas}--\eqref{eq:hatG}, and \eqref{eq:Nrad}--\eqref{eq:bmGprime}, together with the Poisson equation, using an operator-splitting method at each time step.
We decompose the equations into the following five parts and solve them in order.
In what follows, we describe the solution procedure for each step in some detail.
Because the density and temperature of each cell evolve during every substep, the optical depth of a cell also varies accordingly.
Thus, the redistribution of the radiation components based on Equations~\eqref{eq:distribute_ST} and \eqref{eq:distribute_RN} is also performed at every substep in which the effects of radiation are involved.
The detailed procedure for this redistribution is described below.
The treatment of the photon-number density is identical to the radiation energy density and omitted here for brevity.
\subsection{Step 1: hydrodynamics step}
In this step, we solve the following set of equations:
\begin{eqnarray}
    && \f{\p \rho}{\p t} + \nabla\cdot(\rho\bm{v}) = 0 , \label{eq_Step1:rho} \\
    && \f{\p E_\mr{gas}}{\p t} + \nabla\cdot\left\{\bm{v}(E_\mr{gas}+P_\mr{gas})\right\} = 0 , \\
    && \f{\p \rho\bm{v}}{\p t} + \nabla\cdot(\rho\bm{v}\otimes\bm{v}+P_\mr{gas}\bm{I}) = 0 , \label{eq_Step1:rhov} \\
    && \f{\p E^\mr{con}_\mr{rad}}{\p t} + \nabla\cdot\left\{\f{4\bm{v}}{3}(\EofRSLA_\mr{rad,T}+\EofNonRSLA_\mr{rad})\right\} = 0 . \label{eq_Step1:Erad}
\end{eqnarray}
Here, we evaluate the intercell numerical fluxes in Equations~\eqref{eq_Step1:rho}--\eqref{eq_Step1:rhov} using the HLLC approximate Riemann solver \citep{Toro_et_al_1994}.
On the other hand, we compute the fluxes in Equation~\eqref{eq_Step1:Erad} as the product of the interface velocity obtained from the Riemann solver and the $(4/3)(\EofRSLA_\mr{rad,T}+\EofNonRSLA_\mr{rad})$ in the upwind cell.
Here, we estimate $\EofRSLA_\mr{rad,T}$ and $\EofNonRSLA_\mr{rad}$ using the local $\tau_\mr{cell}$ with Equations~\eqref{eq:distribute_ST} and \eqref{eq:distribute_RN}.
\par
In the original SFUMATO code, a monotone upstream-centered scheme for conservation laws and a predictor-corrector method are employed to achieve second-order accuracy in both space and time \citep{Matsumoto_2007}.
However, in regions with steep gradients in physical quantities, such as protostellar surfaces, this type of interpolation can yield inaccurate results and lead to a breakdown of simulations.
To avoid such issues and enhance numerical stability, we adopt a simpler approach with first-order accuracy in both space and time in this study. 
We estimate the values at cell interfaces by taking them equal to those at cell centers, and solve the time evolution using the Euler method.
\subsection{Step 2: gravity step} \label{subsec:gravity_step}
In this step, we solve the Poisson equation and compute the gravitational acceleration:
\begin{eqnarray}
    && \nabla^2 \phi = 4 \pi G \rho , \label{eq:Poisson} \\
    && \f{\p E_\mr{gas}}{\p t} = \rho\bm{g}\cdot\bm{v} , \label{eq:Grav_Egas}\\
    && \f{\p \rho\bm{v}}{\p t} = \rho\bm{g} , \label{eq:Grav_rhov}
\end{eqnarray}
where $G$ is the gravitational constant, $\phi$ is the gravitational potential, and $\bm{g}=-\nabla \phi$ is the gravitational acceleration.
We solve Equation~\eqref{eq:Poisson} to obtain the gravitational potential $\phi$ using a multigrid method, as described in detail by \citet{Matsumoto_2007}.
We then solve Equations~\eqref{eq:Grav_Egas} and \eqref{eq:Grav_rhov} following the method by \citet{Mullen_et_al_2021}.
Unlike conventional self-gravity updates, which treat gravity simply as source terms, their method evaluates the source terms in a way that is mathematically equivalent to flux divergences.
This guarantees conservation of total energy and momentum to round off accuracy.
Moreover, a key advantage of their formulation is that, unlike previous energy-conservative approaches, it also guarantees curl-free gravity, $\nabla \times \bm{g} = 0$.
In simulations of self-gravitating systems, numerical errors that violate either energy–momentum conservation or the curl-free nature of gravity can accumulate and lead to unphysical behavior \citep{Jiang_and_Goodman_2011,Jiang_et_al_2013,Mullen_et_al_2021}.
Therefore, the energy-conservative self-gravity scheme by \citet{Mullen_et_al_2021} is essential for maintaining long-term stability and suppressing spurious numerical artifacts.
We examine in Appendix~\ref{subsec:Com_Self_Gravity} how the protostellar evolution during the collapse of a spherically symmetric gas depends on the choice of self-gravity scheme.
\par
In the method by \citet{Mullen_et_al_2021}, the right-hand side of Equation~\eqref{eq:Grav_Egas} is evaluated with the mass flux obtained from the Riemann solver.
This treatment ensures that the energy source term is consistent with mass conservation.
Following this prescription, we compute the right-hand side of Equation~\eqref{eq:Grav_Egas} using the Riemann mass flux obtained in Step 1 (hydrodynamics step).
Although their implementation employs a predictor–corrector time integrator, we instead use a first-order time integration with the Euler method to enhance numerical stability.
\subsection{Step 3: emission and absorption step} \label{subsec:Step3}
In this step, we solve the energy exchange between the gas and radiation through emission and absorption in each cell,
\begin{eqnarray}
    && \f{\p E_\mr{gas}}{\p t}  =  cG^0, \label{eq:Emit_and_Abs_Egas} \\
    && \f{\p E^\mr{con}_\mr{rad}}{\p t} = -cG^0, \\
    && G^0 = \rho \left\{  \kappa_{0E} (f_c \EofRSLA_\mr{rad} + \EofNonRSLA_\mr{rad}) - \kappa_\mr{0P} a T^4 \right\} .
\end{eqnarray}
Here, we distribute $E^\mr{con}_\mr{rad}$ into $\EofRSLA_\mr{rad}$ and $\EofNonRSLA_\mr{rad}$ following Equation~\eqref{eq:distribute_RN}.
As described in Section~\ref{subsec:RadPartition}, to accurately capture the escape of radiation from an optically thick cell into an optically thin one, we must take into account information from neighboring cells and estimate $\fdisofNonRSLA$ using the minimum cell optical depth among the local and neighboring cells.
Accordingly, we compute $\fdisofNonRSLA$ and distribute $E^\mr{con}_\mr{rad}$ as
\begin{eqnarray}
    && \tau^{i,j,k}_\mr{cell,eff} = \min_{(p,q,r)\in\{(i,j,k)\}\cup\mathcal{N}(i,j,k)} \tau^{p,q,r}_\mr{cell},  \\
    && \fdisofNonRSLA_\mr{eff} = \exp\left(-\f{2}{3f_c \tau_\mr{cell,eff}}\right) , \\
    && \EofRSLAijk_\mr{rad} = (1-\fdisofNonRSLA_\mr{eff})  E^{\mr{con},i,j,k}_\mr{rad} , \\
    && \EofNonRSLAijk_\mr{rad} = \fdisofNonRSLA_\mr{eff} E^{\mr{con},i,j,k}_\mr{rad} , \label{eq:Emit_and_Abs_distribute}
\end{eqnarray}
where the superscripts $(i,j,k)$ and $(p,q,r)$ denote cell indices in the 3D computational grid, and $\mathcal{N}(i,j,k)$ denotes the set of the six face-adjacent neighboring cells of cell $(i,j,k)$.
The use of the six face-adjacent neighboring cells in evaluating $\tau_\mr{cell,eff}$ is motivated by the RT step (Section~\ref{subsec:Step4}), in which radiative transport is solved exclusively between a cell and its six neighboring cells.
If any of these neighboring cells is optically thin, radiation should preferentially escape in that direction.
By defining $\tau_\mr{cell,eff}$ as above, we ensure that sufficient radiation is assigned to the 
RSLA-streaming component when an optically thin escape channel exists.
\par
Since the $G^0$ term is generally very stiff, particularly in optically thick regions, explicit integration would require prohibitively small time steps for numerical stability.
We thus solve Equations~\eqref{eq:Emit_and_Abs_Egas}--\eqref{eq:Emit_and_Abs_distribute} using an implicit method.
However, since $\kappa_{0E}$, $\kappa_{0P}$, and $\tau_\mr{cell,eff}$ depend on $E^\mr{con}_\mr{rad}$, and $T$, the system is nonlinear and cannot be solved analytically.
We therefore adopt an iterative method to obtain the solution.
In this procedure, $\tau_\mr{cell,eff}$ is fixed at its value before the update, since it depends on neighboring cells and updating it self-consistently within the iterative solver would be computationally expensive.
\subsection{Step 4: radiative transfer step} \label{subsec:Step4}
In this step, we solve the RT using the M1 closure scheme for the RSLA component and the diffusion scheme for the non-RSLA component,
\begin{eqnarray}
    && \f{\p E^\mr{con}_\mr{rad}}{\p t} + \nabla\cdot\bra{\bm{F}_\mr{S}-\f{c}{3\hat{\chi}\rho}\nabla \EofNonRSLA_\mr{rad}} = 0 , \label{eq:RTStep_E} \\
    && \f{\p \bm{F}_\mr{S}}{\p t} + \tilde{c}^2\nabla\cdot\PofRSLA_\mr{rad,S} = 0 \label{eq:RTStep_F} .
\end{eqnarray}
For the flux of the RSLA component ($F_\mr{S}$ in Equation~\ref{eq:RTStep_E} and $\PofRSLA_\mr{rad,S}$ in Equation~\ref{eq:RTStep_F}), we estimate the intercell numerical flux with the GLF solver as
\begin{eqnarray}
    F_{\f{1}{2}} = \f{F^\mr{R}+F^\mr{L}}{2} - \f{\tilde{c}}{2} (E^\mr{R} - E^\mr{L}) , \label{eq:GLF}
\end{eqnarray}
where the superscripts $\mathrm{R}$ and $\mathrm{L}$ denote the values of physical quantities in the right and left cells adjacent to the interface, respectively.
As explained in Section~\ref{subsec:RT15}, we need to solve the RT of the RSLA component using only the streaming photons $\EofRSLA_\mr{rad,S}$ to reproduce the physical diffusion by the numerical diffusion represented by the second term in Equation~\eqref{eq:GLF}. 
Therefore, the variables $(E,F)$ in Equation~\eqref{eq:GLF} correspond to ($\EofRSLA_\mathrm{rad,S},\bm{F}_\mathrm{S}$) and ($\bm{F}_\mathrm{S},\PofRSLA_\mr{rad,S}$) for solving Equations~\eqref{eq:RTStep_E} and \eqref{eq:RTStep_F}.
For the flux of the non-RSLA component, $-(c/3\hat{\chi}\rho)\nabla\EofNonRSLA_\mr{rad}$ in Equation~\eqref{eq:RTStep_E}, we compute the intercell numerical flux with the central difference method as
\begin{eqnarray}
    && F_{\f{1}{2}} = - \f{c}{3\{(\hat{\chi}^\mr{R}\rho^\mr{R}+\hat{\chi}^\mr{L}\rho^\mr{L})/2\}} \f{\EofNonRSLAR_\mr{rad} - \EofNonRSLAL_\mr{rad}}{\Delta x}.
\end{eqnarray}
\par
Here, we estimate the value of $\EofRSLA_\mr{rad,S}$ and $\EofNonRSLA_\mr{rad}$ by distributing $E^\mr{con}_\mr{rad}$ based on Equations~\eqref{eq:distribute_ST} and \eqref{eq:distribute_RN}.
However, as mentioned in Section~\ref{subsec:RadPartition}, to capture the physical behavior of radiation escaping from an optically thick cell into an optically thin one, we must ensure that the streaming photons $\EofRSLA_\mr{rad,S}$ dominate even in an optically thick cell if its neighboring cell is optically thin.
To achieve this, we obtain $\EofRSLA_\mr{rad,S}$ and $\EofNonRSLA_\mr{rad}$ using the smaller of the optical depths of the two adjacent cells as
\begin{eqnarray}
    && \tau_\mr{cell,\f{1}{2}} = \min(\tau_\mr{cell}^\mr{R},\tau_\mr{cell}^\mr{L}) , \\
    && \fdis_\mr{T,\f{1}{2}} = \exp\left(-\frac{2}{3\tau_\mr{cell,\f{1}{2}}}\right), \\
    && \fdisofNonRSLA_\mr{\f{1}{2}} = \exp\left(-\f{2}{3f_c \tau_\mr{cell,\f{1}{2}}}\right) , \\
    && \EofRSLARorL_\mr{rad,S} = (1 - \fdis_\mr{T,\f{1}{2}})(1 - \fdisofNonRSLA_{\f{1}{2}}) E_\mr{rad}^\mr{con,\,R\,or\,L} , \\
    && \EofNonRSLARorL_\mr{rad} = \fdisofNonRSLA_{\f{1}{2}} E^\mr{con,\,R\,or\,L}_\mr{rad} .
\end{eqnarray}
This scheme allows radiation to freely escape in directions where optically thin cells are present.
\subsection{Step 5: radiative momentum exchange step}
In the final step, we solve the momentum exchange by the radiative source terms:
\begin{eqnarray}
    && \f{\p \rho\bm{v}}{\p t} = \bmGofRSLA_\mr{S} + \bmGofRSLA_\mr{T} + \bmGofNonRSLA , \\
    && \f{\p \bm{F}_\mr{S}}{\p t} = -c\tilde{c} \bmGofRSLA_\mr{S} , \\
    && \bmGofRSLA_\mr{S} = \rho \chi_{0F} \f{\bm{F}_\mr{S}}{c} - \rho \chi_{0F}\f{4\bm{v}}{3c} f_c \EofRSLA_\mr{rad,S} , \label{eq_Step5:GS} \\
    && \bmGofRSLA_\mr{T} = -  \f{1}{3} \nabla f_c \EofRSLA_\mathrm{rad,T} , 
    \hspace{5mm} \bmGofNonRSLA = -  \f{1}{3} \nabla \EofNonRSLA_\mathrm{rad}  .
\end{eqnarray}
In this step, we estimate $\EofRSLA_\mr{rad,S}$, $\EofRSLA_\mr{rad,T}$ and $\EofNonRSLA_\mr{rad}$ using the local $\tau_\mr{cell}$ with Equation~\eqref{eq:distribute_ST} and \eqref{eq:distribute_RN}.
The first term in Equation~\eqref{eq_Step5:GS} is very stiff, and explicit integration requires an extremely small time step.
To avoid this problem, we solve this term implicitly, as
\begin{eqnarray}
    \bm{F}^{n+1}_\mr{S} &=& \f{1}{1+\rho\chi_{0F}\tilde{c} \Delta t} \bm{F}^n_\mr{S} , \\
    (\rho \bm{v})^{n+1} &=& (\rho \bm{v})^{n} + \rho\chi_{0F} \f{\bm{F}^{n+1}_\mr{S}}{c} \Delta t ,
\end{eqnarray}
where the superscripts $n$ and $n+1$ denote the current and the updated time steps, respectively.
For other terms, we integrate them explicitly and evaluate the spatial derivatives using central difference schemes.
\section{NUMERICAL TESTS} \label{sec:tests}
In this section, we present a series of numerical tests to validate the newly developed RHD scheme. 
In Section~\ref{subsec:Tests_of_non-reduced}, we validate the non-RSLA component introduced in Section~\ref{subsec:HybridRT}, with respect to energy conservation (Section~\ref{subsec:Tests_for_EnergyConservation}), radiation transport (Sections~\ref{subsec:Diffusion} and \ref{subsec:Advecting_Radiation_Pulse}), radiation pressure (Section~\ref{subsec:Radiation_Pressure_Tube}).
In Section~\ref{subsec:Radiation_Escap}, we demonstrate that the method introduced in Section~\ref{subsec:RadPartition}, which incorporates information about the optical depth of neighboring cells, enables accurate treatment of RT even in regions with steep optical-depth gradients.
In Section~\ref{subsec:Tests_for_RadSpec}, we verify that our spectral reconstruction method, described in Section~\ref{sec:spectral_reconstruction}, can track the evolution of the radiation spectrum consistently with the emitting sources, including its spatial variation.
Moreover, in Section~\ref{subsec:overhead}, we measure the computational time of each computational operation in the test described in Section~\ref{subsec:Tests_for_RadSpec} and summarize the overhead of our new scheme.
\par
In all of the following tests, we set the reduction factor of the RSLA $f_c$ to $10^{-3}$ and CFL number to $0.5$ for both the hydrodynamics and the M1 CFL condition.
We summarize the main parameters of each test in Table~\ref{tab:test_parameter}.
\begin{table*}
  \centering
  \caption{Parameters for the numerical tests. We adopt $f_c=10^{-3}$ and a CFL number of 0.5 for all tests.}
  \label{tab:test_parameter}
  \begin{tabular}{lccccccc}
    \hline
    \multirow{2}{*}{Section} & \multirow{2}{*}{Box Size} & \multirow{2}{*}{Resolution}  & 
    $\kappa_{0}^\nu$ & $\sigma_{0}^\nu$ & \multirow{2}{*}{$\mr{BC}_\mr{hydro}$} & \multirow{2}{*}{$\mr{BC}_\mr{rad}$} \\
    & & & ($\mr{cm}^2~\mr{g}^{-1}$) & ($\mr{cm}^2~\mr{g}^{-1}$) & & \\
    \hline
    \ref{subsec:Tests_for_EnergyConservation} & $10^{11}$~cm & 1 & 10 & 0 & --- & --- \\
    \ref{subsec:Diffusion}   & $200$~au & $32^3$ & 0 & $4.3\times10^{-2}$ & --- & optically thick \\
    \ref{subsec:Advecting_Radiation_Pulse} & $1024$~cm & $256^3$ & $10^4$ & 0 & peridodic & periodic \\
    \ref{subsec:Radiation_Pressure_Tube}$^{\,*1}$ & $128$~cm & $128^3$ & $10^4$ & 0 & symmetric/periodic & Dirichlet/periodic \\
    \ref{subsec:Radiation_Escap} & $200$~au & $128^3$ & $10^2$ & 0 & --- & optically thin \\
    \ref{subsec:Tests_for_RadSpec}$^{\,*2}$ & $320$~au & $256^3$ & $10^2$/$0$ & $0$/$10^{2}$  & --- & optically thick \\
    \hline
    \multicolumn{7}{l}{Notes. $\mr{BC}_\mr{hydro}$ and $\mr{BC}_\mr{rad}$ denote the outer boundary conditions for hydrodynamics and radiation, respectively.} \\
    \multicolumn{7}{l}{The term ``optically thick'' refers to a boundary condition in which the radiation energy density decreases as $r^{-1}$} \\
    \multicolumn{7}{l}{and the flux decreases as $r^{-2}$. The term ``optically thin'' refers to a boundary condition where both the energy density} \\
    \multicolumn{7}{l}{and flux decrease as $r^{-2}$.} \\
    \multicolumn{7}{l}{$^{*1}$ Boundary conditions for $x$/($y$, $z$) directions are denoted.}\\
    \multicolumn{7}{l}{$^{*2}$ Opacity values inside/outside clumps are denoted.}
  \end{tabular}
\end{table*}
\subsection{Tests of the non-RSLA component} \label{subsec:Tests_of_non-reduced}
\subsubsection{Energy conservation test during gas-radiation interaction} \label{subsec:Tests_for_EnergyConservation}
\begin{figure*}[t]
  \begin{center}
    \includegraphics[width=\linewidth]{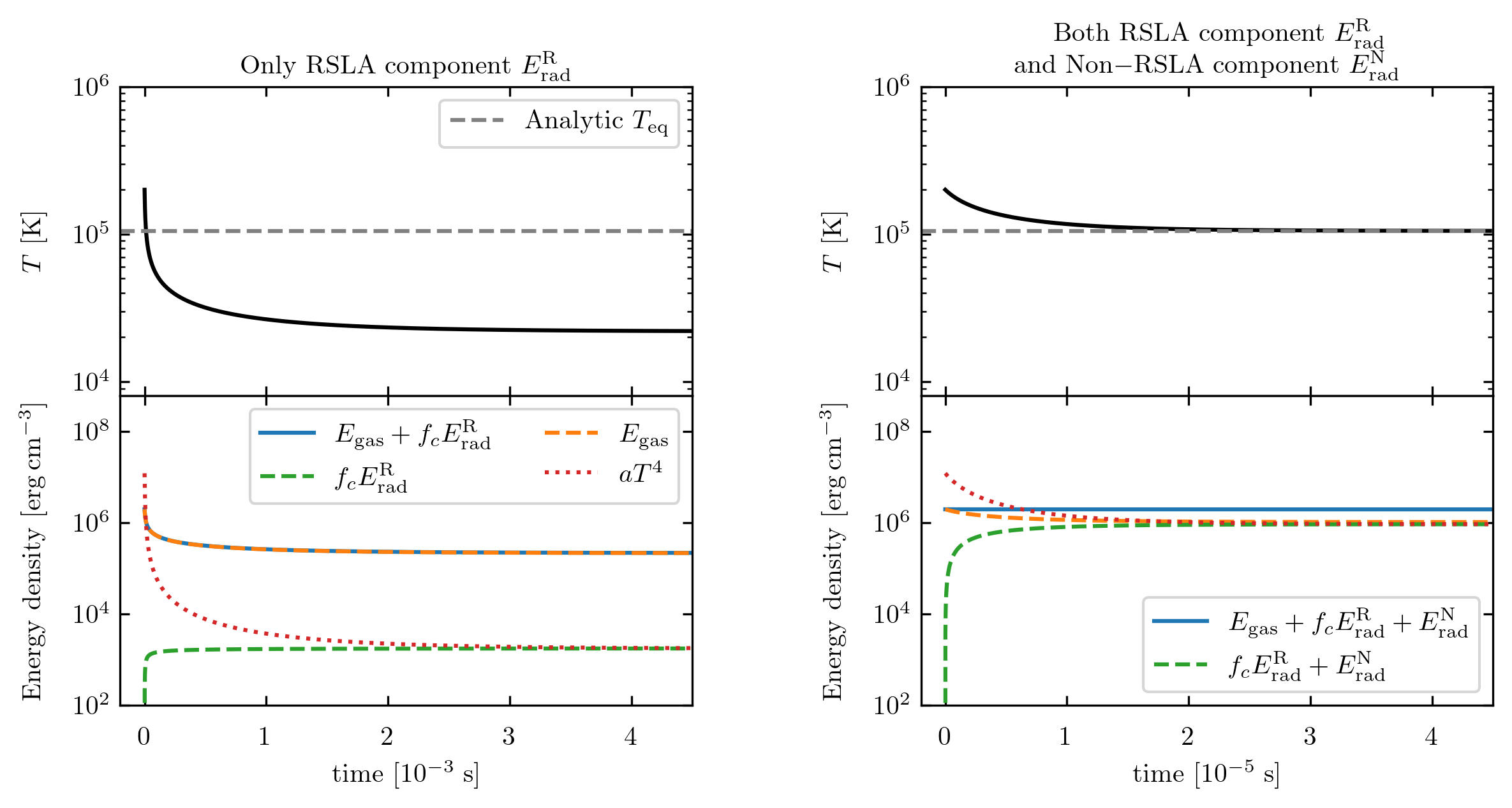}
    \caption{
        Energy conservation test during gas-radiation interaction.  
        The left and right panels show the results without and with the inclusion of the non-RSLA radiation component $\EofNonRSLA_\mathrm{rad}$, respectively.  
        The upper panels display the time evolution of the gas temperature, and the lower panels show the evolution of the energy density of each component.  
        In the upper panels, the black solid lines show the temperature evolution, and the gray-dashed lines indicate the analytic equilibrium temperature when the system evolves while conserving its total energy (see text).
        In the lower panels, the blue solid line represents the total energy (gas plus radiation), the orange- and green-dashed lines denote the gas and radiation energy densities, respectively, and the red-dotted line indicates the radiation energy density corresponding to the Planck distribution, $aT^4$. 
    }
    \label{fig:chemequ_test}
  \end{center}
\end{figure*}
As discussed in Section~\ref{subsec:energy-nonconserve}, the use of the RSLA can lead to violations of energy conservation during the energy exchange between gas and radiation, making it difficult to accurately follow the evolution of optically thick, radiation-dominated systems.
To address this issue, in Section~\ref{subsec:HybridRT}, we have introduced a non-RSLA radiation component, $\EofNonRSLA_\mathrm{rad}$, which is not subject to the RSLA and becomes dominant in optically thick regions such that $\tau_\mr{cell}\gtrsim 2/(3f_c)$.
To verify the effectiveness of this approach, we evolve a single computational cell over time under local emission and absorption processes, without including hydrodynamics or RT.
This test demonstrates that our new scheme conserves the total energy and accurately reproduces the thermal equilibrium state expected under optically thick conditions.
\par
As an initial condition, we consider a gas with a density of $\rho = 10^{-7}~\mathrm{g}~\mathrm{cm}^{-3}$ and a temperature of $2 \times 10^{5}~\mathrm{K}$, and no initial radiation field. 
Through local emission and absorption, the gas loses internal energy while the radiation energy increases, and the system gradually approaches thermal equilibrium characterized by the radiation energy density of $a T^4$. 
We adopt a mean molecular weight of $\mu = 1.27$ and an adiabatic index of $\gamma = 5/3$.
The absorption opacity is set to a constant value of $\kappa^\nu_{0} = 10~\mathrm{cm}^2~\mathrm{g}^{-1}$, and the cell width is fixed at $10^{11}~\mathrm{cm}$, resulting in an optical depth per cell of $\tau_\mathrm{cell} = 10^5$.
Under these conditions, as can be seen from Equations~\eqref{eq:distribute_ST} and \eqref{eq:distribute_RN}, the non-RSLA component $\EofNonRSLA_\mathrm{rad}$ dominates the total radiation energy budget.
\par
When the system evolves while conserving its total energy, the initial gas temperature $T_0$ and the final equilibrium temperature of gas and radiation $T_{\mathrm{eq}}$ should satisfy
\begin{eqnarray}
    \frac{1}{\gamma - 1} \frac{\rho k_B T_0}{\mu m_{\mathrm{H}}}
    = \frac{1}{\gamma - 1} \frac{\rho k_B T_{\mathrm{eq}}}{\mu m_{\mathrm{H}}}
    + a T_{\mathrm{eq}}^4 .
\end{eqnarray}
The left-hand side represents the initial internal energy density of the gas, and the right-hand side corresponds to the sum of the final gas internal energy density and the radiation energy density. 
Solving the above equation under the given conditions yields an analytic equilibrium temperature of $T_{\mathrm{eq}} = 1.05 \times 10^5~\mathrm{K}$.
\par
Figure~\ref{fig:chemequ_test} shows the time evolution of the gas temperature (top panels) and energy density (bottom panels).
The left and right panels correspond to simulations without and with the non-RSLA radiation component $\EofNonRSLA_\mathrm{rad}$, respectively.
It should be noted that, for the RSLA component, the physical energy density is not $\EofRSLA_\mr{rad}$ but $f_c \EofRSLA_\mathrm{rad}$ (see Equation~\ref{eq:def_tildeErad}).
As seen in the lower left panel, in the absence of $\EofNonRSLA_\mathrm{rad}$, the radiation energy density (the green-dashed line) converges to $aT^4$ (the red-dotted line), meaning that the system approaches thermal equilibrium.
However, the total energy density (the blue solid line), i.e., the sum of the gas and radiation energy densities, decreases over time, indicating a violation of energy conservation during the gas-radiation interaction. 
Consequently, the final temperature (the top left panel) falls below the physically expected value represented by the gray-dashed line.
\par
In contrast, the lower right panel shows that, when we introduce the non-RSLA component $\EofNonRSLA_\mathrm{rad}$ in calculations, the system reaches equilibrium while keeping its total energy (the blue line) constant throughout the evolution.
The final radiation energy is more than two orders of magnitude larger than in the case without $\EofNonRSLA_\mathrm{rad}$.
In this case, the final gas temperature approaches the physically expected value as seen in the top right panel.
This result demonstrates that introducing the non-RSLA component $\EofNonRSLA_\mr{rad}$ allows the system to evolve while satisfying energy conservation in the optically thick limit.
\par
Note that the timescale to reach thermal equilibrium differs significantly depending on the presence of $\EofNonRSLA_\mathrm{rad}$. 
In the right panels, the system reaches equilibrium in $\sim 10^{-5}$~s, whereas, in the left panels, it takes $\sim 10^{-3}$~s. 
This is because the gas emissivity, which is proportional to $T^4$, decreases as the temperature drops, leading to a longer time required to reach a lower equilibrium temperature in the left panel.
\subsubsection{Radiative transfer test with the diffusion approximation} \label{subsec:Diffusion}
\begin{figure}[t]
  \begin{center}
    \includegraphics[width=\linewidth]{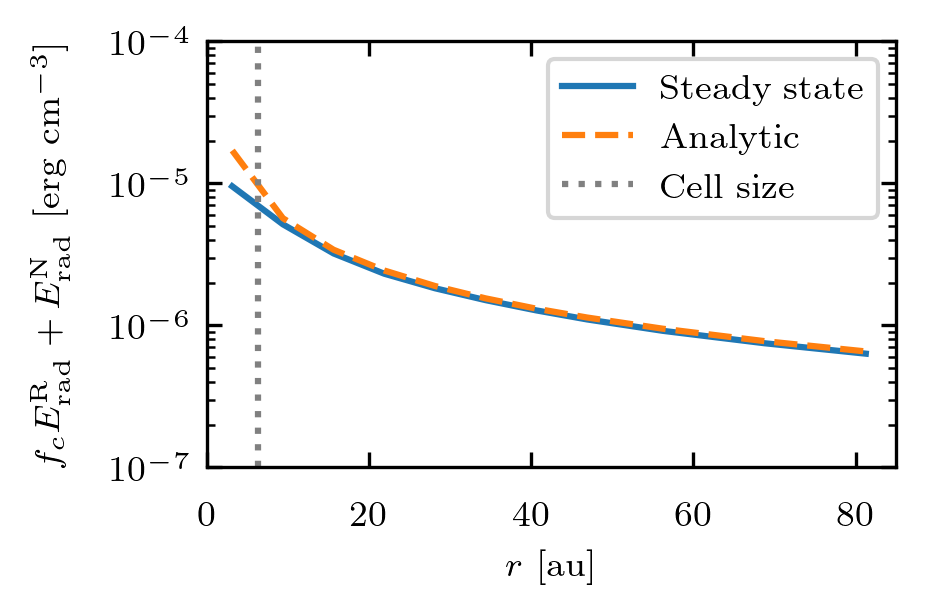}
    \caption{ 
        Radiative transfer test with the diffusion approximation.  
        The blue solid line represents the spherically averaged radial profile in the steady state from our simulation.
        The orange-dashed line denotes the analytic solution.
        The gray-dotted line indicates the computational cell size (6.25~au), which corresponds to the finite volume over which radiation is injected.
        The simulation is performed in a cubic box of size 200 au with a uniform grid of $32^3$.
    }
    \label{fig:transfer_test}
  \end{center}
\end{figure}
In this section, we verify that our scheme can accurately solve the RT of the non-RSLA component $\EofNonRSLA_\mr{rad}$ in the sufficiently optically thick medium.
To isolate the behavior of RT in this test, we turn off hydrodynamics, ignore emission and absorption, and consider only the scattering effect.
We put a radiation source with a luminosity of $10^{30}~\mathrm{erg}~\mathrm{s}^{-1}$ at the center of the computational domain and follow the propagation of radiation through a uniform scattering medium.
The computational box width is $200~\mathrm{au}$, and we resolve the box with $32^3$ cells, resulting in
a cell size of $6.25$~au. 
We adopt a constant gas density of $\rho = 2.3 \times 10^{-9}~\mathrm{g}~\mathrm{cm}^{-3}$ and a scattering opacity of $\sigma^\nu_{0} = 4.3 \times 10^{-2}~\mathrm{cm}^2~\mathrm{g}^{-1}$, corresponding to total optical depths across the box of $\tau_\mathrm{box} = 3 \times 10^{5}$, and to optical depths per cell of $\tau_\mr{cell} = 9.4 \times 10^{3}$.
According to Equations~\eqref{eq:distribute_ST} and \eqref{eq:distribute_RN}, this parameter choice ensures that the dominant radiation component is $\EofNonRSLA_\mathrm{rad}$.
\par
When radiation from a central source propagates through a uniform medium, the steady-state solution can be obtained analytically.
In the optically thick limit, radiation propagates with diffusion as $F = -c \nabla E_\mathrm{rad} / (3 \rho \sigma^\nu_{0})$.
On the other hand, assuming a steady state, the flux can be expressed with the luminosity of the central radiation source $L$ as $F = L / (4\pi r^2)$. As mentioned in \citet{Rosdahl_and_Teyssier_2015}, combining the two expressions for the flux yields the radiation energy density of
\begin{eqnarray}
E_\mathrm{rad} = \frac{3 \rho \sigma^\nu_{0} L}{4 \pi c r} .
\end{eqnarray}
Although this analytic solution assumes an infinite medium, our simulations have a finite computational domain. 
Therefore, we need to impose appropriate outer boundary conditions to reproduce the analytic solution.
To this end, we set the outer boundary conditions of the radiation energy density and flux to match the analytic gradients, given by
\begin{eqnarray}
  U_\mr{0} = U_1 \bra{\f{r_1}{r_0}}^\alpha , \label{eq:BC_U}
\end{eqnarray}
where $U$ represents either the radiation energy density or flux, and the subscripts $0$ and $1$ refer to the ghost cell and the boundary cell inside the computational domain, respectively. 
We set $\alpha=1$ for the energy density and $\alpha=2$ for the flux. 
For all other variables, we adopt free boundary conditions.
\par
Figure~\ref{fig:transfer_test} compares the steady-state radiation energy density $f_c \EofRSLA_\mathrm{rad} + \EofNonRSLA_\mr{rad}$ obtained from our simulation (the blue solid line) with the corresponding analytic solution (the orange-dashed line).
For the simulation result, we present a spherically averaged radial profile.
Except near the center, the radiation energy density obtained from the simulation decreases as $r^{-1}$ in good agreement with the analytic solution.
The small discrepancy near the center arises because the analytic solution assumes an infinitesimal point source, whereas, in the simulation, radiation is injected over a finite volume corresponding to the cell size of $6.25$~au marked by the gray-dotted line.
As we increase the resolution, the region where this discrepancy appears becomes smaller.
This result demonstrates that we can accurately solve the radiative diffusion when the non-RSLA component $\EofNonRSLA_\mr{rad}$ dominates.
\subsubsection{Advecting radiation pulse test} \label{subsec:Advecting_Radiation_Pulse}
\begin{figure}[t]
  \begin{center}
    \includegraphics[width=\linewidth]{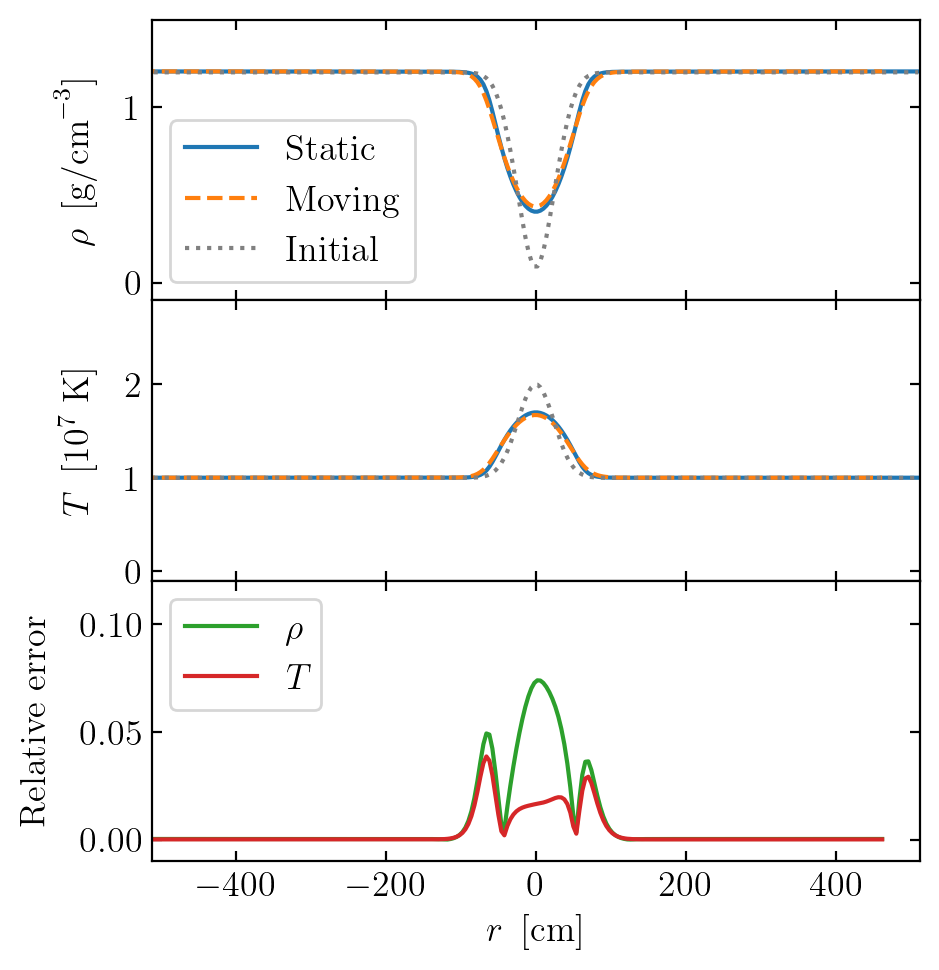}
    \caption{
        Advecting radiation pulse test.  
        The top and middle panels show the distributions of density and temperature.
        The blue solid and orange-dashed lines represent the static and moving cases, respectively, while the gray-dotted lines denote the initial conditions.
        The bottom panel presents the relative errors between the static and moving cases.
        The green and red lines indicate the relative errors in density and temperature, respectively.
        The simulation is performed in a cubic box of size 1024 cm with a uniform grid of $256^3$.
        All quantities are evaluated at $t = 4.8 \times 10^{-4}$~s.
    }
    \label{fig:advection_radpulse}
  \end{center}
\end{figure}
In extremely optically thick environments, such as protostellar interiors, the so-called dynamic diffusion limit applies, in which radiation is transported more efficiently via advection with the moving gas than by diffusion.
In this section, we test whether our code can accurately capture this advection.
As an initial condition, we set a gas distribution in pressure equilibrium, featuring a centrally enhanced radiation energy density accompanied by a corresponding decrease in gas density:
\begin{eqnarray}
  && T = T_0 + (T_1 - T_0) \exp\bra{-\f{x^2}{2w^2}} , \\
  && \rho = \rho_0 \f{T_0}{T} + \f{a\mu m_\mr{H}}{3k_\mr{B}}\bra{\f{T_0^4}{T}-T^3} ,
\end{eqnarray}
where $T_0 = 10^7$~K, $T_1 = 2\times10^7$~K, $\rho_0 = 1.2~\mr{g~cm}^{-3}$, $w = 24$~cm, and $\mu = 2.23$.
The resulting initial profiles are shown as gray-dotted lines in the top and middle panels of Figure~\ref{fig:advection_radpulse}.
As the system evolves, the central radiation peak gradually broadens due to diffusion.
We perform this test under two conditions: one in which the gas remains static, and another in which the gas moves uniformly at a constant velocity of $10^5~\mathrm{cm~s^{-1}}$.
If the scheme accurately captures radiation advection in the dynamic diffusion limit, both cases should exhibit identical evolution in the comoving frame \citep{Krumholz_et_al_2007_Algorithms}.
We set the computational domain to be a cubic box with a side length of $1024$~cm, resolved by $256^3$ uniform cells.
Periodic boundary conditions are applied in all directions.
We adopt a constant absorption opacity of $\kappa^\nu_{0} = 10^4~\mathrm{cm}^2~\mathrm{g}^{-1}$ and neglect scattering.
Under these conditions, the optical depth per cell is $\tau_\mathrm{cell} \gtrsim 4 \times 10^3$, indicating that the radiation energy is dominated by the non-RSLA component $\EofNonRSLA_\mathrm{rad}$.
\par
Figure~\ref{fig:advection_radpulse} compares the profiles obtained in the static and moving cases after evolving the system for $4.8 \times 10^{-4}$~s, during which the moving gas travels a distance of $48$~cm.
To facilitate comparison, we shift the radiation pulse in the moving case by $48$~cm so that its center aligns with that of the static case.
In the top and middle panels, we show the density and temperature profiles for the static case and moving cases, represented by the blue solid and orange-dashed lines, respectively.
As shown in the figure, the two cases exhibit nearly identical evolution, in which radiation diffuses out from the central region, reducing the radiation pressure and allowing gas from the surroundings to flow inward, leading to an increase in the central density.
The bottom panel shows the relative errors, defined as the absolute difference between the two cases, normalized by the value of the static case.
The relative errors in both density (the green line) and temperature (the red line) remain within 8\%.
These results confirm that our scheme accurately captures the advection of radiation by the moving gas in the dynamic diffusion limit.
\subsubsection{Radiation-pressure tube test} \label{subsec:Radiation_Pressure_Tube}
\begin{figure}[t]
  \begin{center}
    \includegraphics[width=\linewidth]{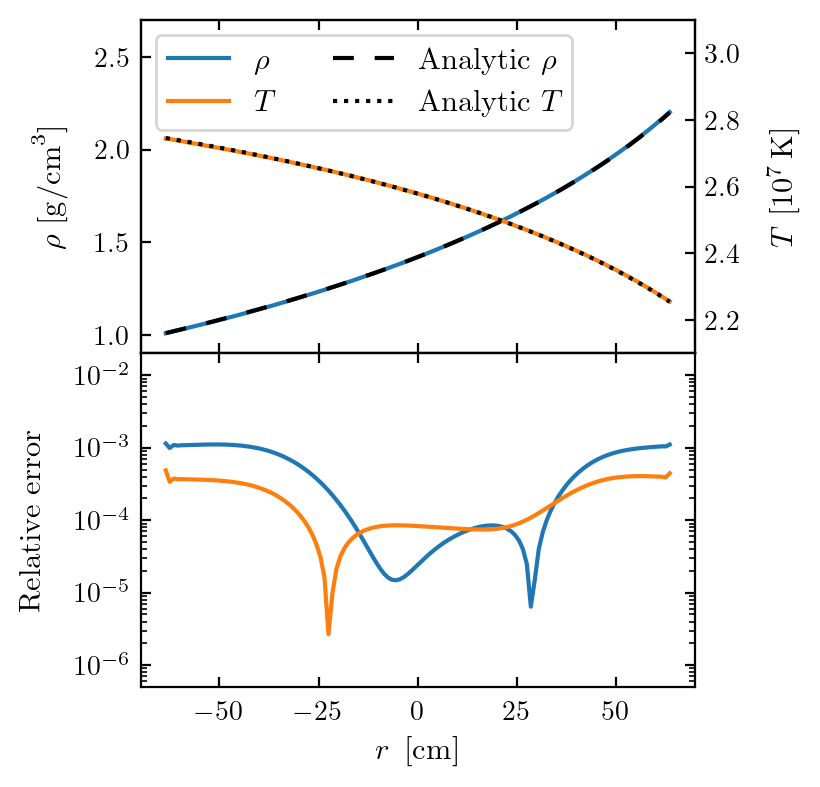}
    \caption{
        Radiation-pressure tube test.
        In the top panel, the blue and orange lines show the density $\rho$ and temperature $T$ profiles along the $x$-axis after ten sound-crossing times, while the dashed and dotted lines represent the analytic hydrostatic equilibrium solutions for density and temperature, respectively.
        In the bottom panel, we show the relative errors of the numerical results presented in the top panel with respect to the analytic solutions. The blue and orange lines represent the relative errors in density and temperature, respectively.
        The simulation is performed in a cubic box of size 128 cm with a uniform grid of $128^3$.
    }
    \label{fig:radpressure_tube}
  \end{center}
\end{figure}
Here, we demonstrate that our scheme accurately captures radiation pressure in the optically thick regime, where the non-RSLA component $\EofNonRSLA_\mathrm{rad}$ dominates. 
To this end, we examine whether our scheme can maintain a hydrostatic structure supported by both gas and radiation pressure.
First, to construct the initial condition for 3D simulations, we obtain the gas profile in steady-state hydrostatic equilibrium under optically thick conditions, where radiation is transported by diffusion, by solving the following 1D equations, which describe the so-called radiation-pressure tube problem \citep{Krumholz_et_al_2007_Algorithms}:
\begin{eqnarray}
  && \bra{\f{k_\mr{B}}{\mu}\rho+\f{4}{3}aT^3}\f{dT}{dx}+\f{k_\mr{B}}{\mu}T\f{d\rho}{dx}=0 , \\
  && \f{d^2T}{dx^2}+3\f{1}{T}\bra{\f{dT}{dx}}^2 - \f{1}{\rho} \bra{\f{d\rho}{dx}} \bra{\f{dT}{dx}} = 0 . 
\end{eqnarray}
The first equation represents the momentum balance between gas pressure and radiation pressure, while the second ensures a constant radiative flux under the diffusion approximation.
We solve these equations by integrating from $x_0 = -64~\mathrm{cm}$ to $x_1 = 64~\mathrm{cm}$, assuming a mean molecular weight of $\mu = 2.33$.
At the boundary $x_0$, we impose $(d\rho/dx)_0 = 5 \times 10^{-3}~\mathrm{g~cm}^{-4}$ and $T_0 = 2.75 \times 10^7~\mathrm{K}$.
For constructing the initial condition for 3D simulation, we map the resulting 1D solution along the $x$-axis onto a 3D computational domain with a side length of $128~\mathrm{cm}$ and a resolution of $128^3$ cells.
Furthermore, we set the physical quantities to be uniform in the $y$- and $z$-directions. 
Along the $x$-axis, Dirichlet boundary conditions are applied to the radiation fields by fixing their values to the exact solution, while symmetry boundary conditions are imposed on the hydrodynamic variables to ensure that the total gas mass within the domain remains constant.
Periodic boundary conditions are applied along the $y$- and $z$-directions.
We perform calculations for a constant absorption opacity of $10^{4}~\mathrm{cm}^2~\mathrm{g}^{-1}$, and neglect scattering.
This parameter choice corresponds to cell optical depths of $\tau_\mr{cell}=10^5$--$2\times10^5$, representing regimes where $\EofNonRSLA_\mathrm{rad}$ dominates.
\par
Figure~\ref{fig:radpressure_tube} presents the simulated gas profiles along the $x$-axis after 10 sound-crossing times and their relative errors with respect to the exact hydrostatic solution.
In the top panel, the blue and orange lines show the density and temperature profiles, respectively, and the dashed and dotted lines denote the corresponding analytic solutions.
This panel indicates the simulation results remain in good agreement with the reference solutions even after 10 sound-crossing times.
The bottom panel displays the relative errors, defined as the absolute difference between the numerical and exact values divided by the exact value.
The errors remain within approximately $0.1$~\% throughout the computational domain.
These results demonstrate that our scheme accurately captures the radiation-pressure force and maintains hydrostatic equilibrium when the non-RSLA component $\EofNonRSLA_\mr{rad}$ dominates.
\subsection{Radiative transfer test with steep optical-depth gradient}  \label{subsec:Radiation_Escap}
\begin{figure}[t]
  \begin{center}
    \includegraphics[width=\linewidth]{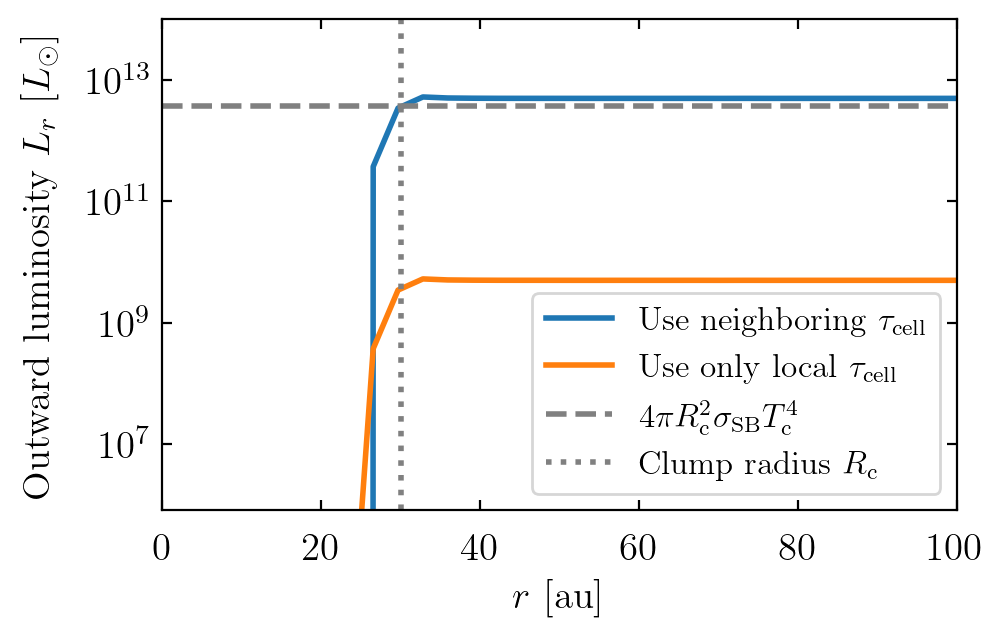}
    \caption{
        Radiative transfer test with steep optical-depth gradient. 
        We show spherically averaged radial profiles of outward luminosity in the steady state.  
        The blue line represents the result obtained by accounting for the cell optical depth $\tau_\mathrm{cell}$ of adjacent cells, while the orange line shows the result obtained using only the local value of $\tau_\mathrm{cell}$.  
        The horizontal dashed line marks the expected luminosity, $4\pi R_\mathrm{c}^2 \sigma_\mathrm{SB} T_\mathrm{c}^4$.
        The vertical dotted line indicates the clump radius, $R_\mathrm{c}$. 
        The simulation is performed in a cubic box of size $200$~au with a uniform grid of $128^3$.
    }
    \label{fig:rad_escape}
  \end{center}
\end{figure}
As mentioned in Section~\ref{subsec:RadPartition}, if radiation is divided into the RSLA and non-RSLA components solely based on the local value of $\tau_\mathrm{cell}$, it cannot escape from optically thick cells even when the adjacent cell is optically thin.
To address this issue, we distribute the radiation components by also considering the $\tau_\mathrm{cell}$ values of neighboring cells, as detailed in Section~\ref{subsec:Step3} and \ref{subsec:Step4}.
In this section, we verify that this method allows radiation to escape from optically thick regions into adjacent optically thin regions.
To isolate the radiative behavior, we disable hydrodynamics and neglect scattering for simplicity.
\par
We set up a computational box with a side length of $200~\mathrm{au}$, resolved with $128^3$ uniform cells.
At the center of the box, we place a gas clump with the radius $R_\mathrm{c} = 30~\mathrm{au}$, density $\rho_\mathrm{c} = 10^{-2}~\mr{g}~\mr{cm}^{-3}$, and temperature $T_\mathrm{c} = 10^5~\mathrm{K}$.
Outside the clump, we place a gas with a density of $\rho = 10^{-20}~\mathrm{g}~\mathrm{cm}^{-3}$ and an initial temperature of $T = 100~\mathrm{K}$.
The gas temperature within the clump is kept fixed throughout the calculation.
We set a constant absorption opacity of $\kappa^\nu_{0} = 10^2~\mathrm{cm^2~g^{-1}}$.
As the outer boundary condition for radiation, we adopt Equation~\eqref{eq:BC_U} with $\alpha = -2$, corresponding to an optically thin radiation field whose intensity decreases as $r^{-2}$.
Due to this opacity value, the optical depth per cell becomes extremely large inside the clump, reaching $\tau_\mathrm{cell} = 5\times10^{11}$, while the region outside the clump remains optically thin.
We then follow the radiation emitted by the clump.
In the ideal case, the clump radiates from its surface with a luminosity of $L = 4\pi R_\mathrm{c}^2 \sigma_\mathrm{SB} T_\mathrm{c}^4$.
\par
Figure~\ref{fig:rad_escape} shows the spherically averaged radial profile of the outward luminosity in the steady state.
Here, the outward luminosity is defined as $L_r = 4 \pi r^2 (F_{\mr{S},r} + F_{\mr{num},r} + \FofNonRSLA_{\mr{diff},r})$, where $F_{\mr{S},r}$ is the radial component of $F_{\mr{S}}$, and $F_{\mr{num},r}$ and $\FofNonRSLA_{\mr{diff},r}$ are the radial components of the flux given by Equations~\eqref{eq:Fnum} and \eqref{eq:Fhat_diff}.
We compare two cases in this figure: one in which the radiation partitioning accounts for the optical depths $\tau_\mathrm{cell}$ of neighboring cells (the blue line), and another that uses only the local value of $\tau_\mathrm{cell}$ (the orange line). 
The dashed horizontal line represents the expected clump luminosity under the current settings, given by $L = 4\pi R_\mathrm{c}^2 \sigma_\mathrm{SB} T_\mathrm{c}^4$, while the dotted vertical line indicates the clump radius, $R_\mathrm{c} = 30$~au.
In both cases, the radiative flux is nearly zero inside the clump, while the luminosity increases sharply at the clump surface. 
In the optically thin regions outside the clump, the luminosity remains constant as radiation propagates outward.
When only the local $\tau_\mathrm{cell}$ is considered, the resulting luminosity falls significantly short of the expected value of $4\pi R_\mathrm{c}^2 \sigma_\mathrm{SB} T_\mathrm{c}^4$.
This is because, under the current setup, the entire radiation energy is assigned to the non-RSLA component $\EofNonRSLA_\mathrm{rad}$ within the clump, thereby suppressing the radiation escape from the clump surface.
In contrast, in our improved scheme, the radiation energy in the outermost cell of the clump is assigned to the streaming component $\EofRSLA_\mathrm{S}$, allowing the physically expected luminosity to be reproduced with good accuracy.
In addition, although the above test assumes that the non-RSLA component $\EofNonRSLA_{\mathrm{rad}}$ dominates inside the clump, our scheme also works well when the trapped RSLA component $\EofRSLA_{\mathrm{rad,T}}$ is dominant inside the clump.
These results demonstrate the validity of our method.
\subsection{Radiation spectrum test} \label{subsec:Tests_for_RadSpec}
\begin{figure}[t]
  \begin{center}
    \includegraphics[width=\linewidth]{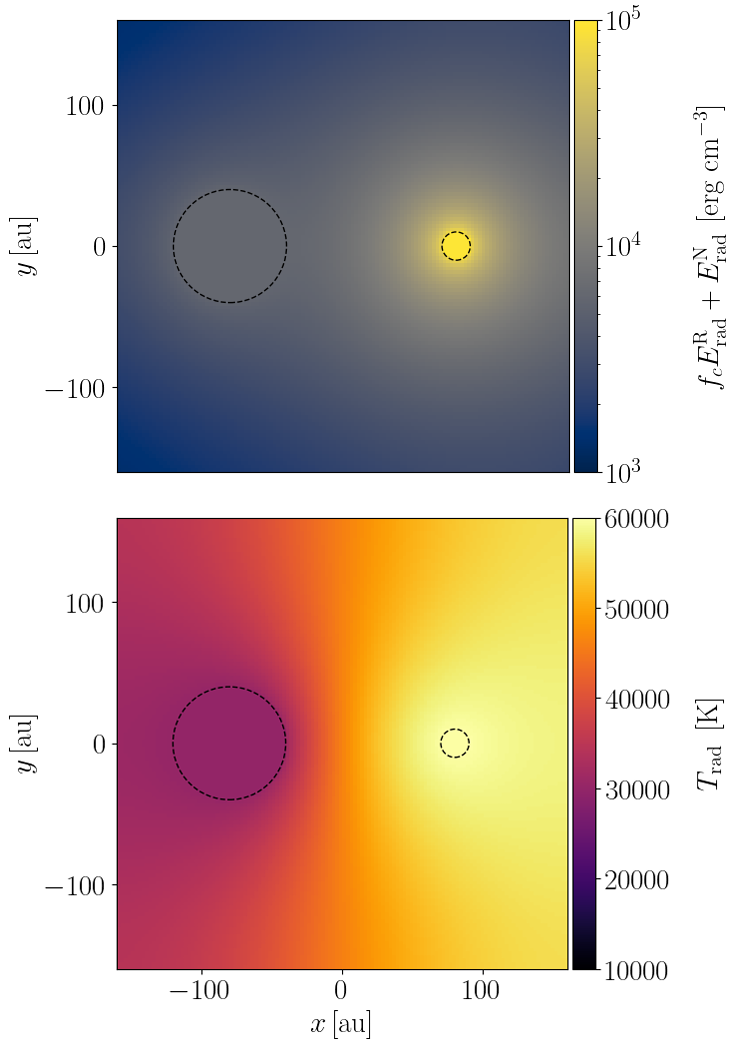}
    \caption{
        Radiation spectrum test. The top and bottom panels show the distributions of radiation energy density and radiation temperature, respectively, in the steady state.
        These panels present an $xy$-slice at $z = 0$ from the 3D simulation. Dashed lines indicate the surfaces of the clumps, within which the gas temperature is fixed.
        The simulation is performed in a cubic box of size $320$~au with a uniform grid of $256^3$.
    }
    \label{fig:Spectrum}
  \end{center}
\end{figure}

In our scheme, the radiation spectrum is reconstructed from the photon mean energy, as described in Section~\ref{sec:spectral_reconstruction}.
In this section, we validate the ability of our method to accurately track the evolution of the radiation spectrum. 
We set up a computational box with a side length of $320~\mathrm{au}$, resolved with $256^3$ uniform cells.
Then, as illustrated by the dashed line in Figure~\ref{fig:Spectrum}, we place two clumps along the $x$-axis at positions $x = 80$ and $-80$~au, with temperatures of $T_\mr{c}=6 \times 10^4$ and $3 \times 10^4$~K, and radii of $R_\mr{c} =10$ and $40$~au, respectively.
Both clumps have identical luminosities, given by $L = 4\pi R_\mathrm{c}^2 \sigma_\mathrm{SB} T_\mathrm{c}^4$.
The gas density inside each clump is set to $10^{-2}~\mathrm{g~cm}^{-3}$, while the ambient medium has a density of $10^{-16}~\mr{g~cm}^{-3}$ and an initial temperature of $100$~K.
To isolate radiation effects, we turn off hydrodynamics and fix the gas temperature within the clumps.
The opacity is set such that $\kappa^\nu_{0}=100~\mr{cm}^2~\mr{g}^{-1}$ and $\sigma^\nu_{0}=0~\mr{cm}^2~\mr{g}^{-1}$ inside the clumps, and $\kappa^\nu_{0}=0~\mr{cm}^2~\mr{g}^{-1}$ and $\sigma^\nu_{0}=100~\mr{cm}^2~\mr{g}^{-1}$ outside.
This setup ensures that radiation is emitted only within the clumps, while outside it is scattered without absorption or emission. 
As the outer boundary condition for radiation, we adopt the same condition as in Section~\ref{subsec:Diffusion}, which is appropriate for an optically thick radiation field.
Starting from an initial condition with no radiation field, we evolve the system until it reaches a steady state.
\par
Figure~\ref{fig:Spectrum} shows a 2D snapshot on the $xy$-plane at $z=0$ in the steady state.
The top panel presents the radiation energy density $f_c \EofRSLA_\mr{rad} + \EofNonRSLA_\mr{rad}$, which peaks at the positions of the clumps and gradually decreases with distance. 
The clump on the right exhibits a higher radiation energy density due to its higher temperature.
The bottom panel displays the radiation temperature $T_\mr{rad}$ calculated according to Equation~\eqref{eq:get_Trad}.
Each clump emits radiation at a temperature equal to its own gas temperature, and in the surrounding regions, $T_\mathrm{rad}$ reflects that of the nearest clump.
In the region between the two clumps, the radiation temperature takes an intermediate value between the two source temperatures.
This test demonstrates that our scheme accurately captures the spatial variation of the radiation temperature from multiple sources, thereby enabling a consistent reconstruction of the radiation spectrum.
\par
\subsection{Computational overhead of RT} \label{subsec:overhead}
We comment on the increase in wall-time of RT introduced by the newly implemented scheme.
We measure the wall-time 
fractions for the test calculation in Section~\ref{subsec:Tests_for_RadSpec}, 
as summarized in Table~\ref{tab:computational_overhead}.
The computational cost increases due to the inclusion of the non-RSLA component. 
However, because this diffusive RT involves a simpler flux evaluation than the M1 RT,
its associated overhead is limited to $8\%$, compared to $41\%$ for the M1 RT.
The evaluation of $\tau_\mathrm{cell}$, which 
requires collecting information from neighboring cells, is a relatively simple operation and accounts for only $1\%$ of the total wall-time.
Evolving the photon-number density accounts for $50\%$ of the total RT wall-time, effectively doubling the computational cost of the RT solver,
because the procedures for evolving the photon-number density are similar to those for evolving the photon energy density.
Overall, the total computational cost of the newly implemented scheme is approximately a factor of $100/41\simeq2.4$ higher than that of an M1 RT scheme evolving only the radiation energy density.
\begin{table}
    \centering
    \caption{Wall-time percentages for each computational operation in the radiation spectrum test in Section~\ref{subsec:Tests_for_RadSpec}.
    }
    \label{tab:computational_overhead}
    \begin{tabular}{lcc}
        \hline
        Operation & Percentage (\%) \\
        \hline
        M1 RT of RSLA component   & 41  \\
        Diffusive RT of Non-RSLA component  & 8  \\
        Estimation of $\tau_\mr{cell}$ & 1 \\
        RT of photon-number density  & 50 \\
        \hline
    \end{tabular}
\end{table}
\section{Astrophysical Application: 3D Simulation of Protostar Evolution in the Early Universe} \label{sec:ProsForm}
In this section, we demonstrate an astrophysical application of the RHD code developed in this study. 
Specifically, we simulate the evolution of a protostar forming in the early Universe under extremely high accretion rates, corresponding to the so-called direct collapse scenario \citep{Bromm_and_Loeb_2003, Begelman_et_al_06, Inayoshi_et_al_2014}.
While a similar problem was previously investigated in \citet{Kimura_et_al_2023}, where an artificial Bonnor–Ebert sphere was adopted as the initial condition, here, we extend the study by starting from cosmological initial conditions, thereby providing a more self-consistent treatment of the early formation environment.
\begin{figure*}[t]
  \begin{center}
    \includegraphics[width=\linewidth]{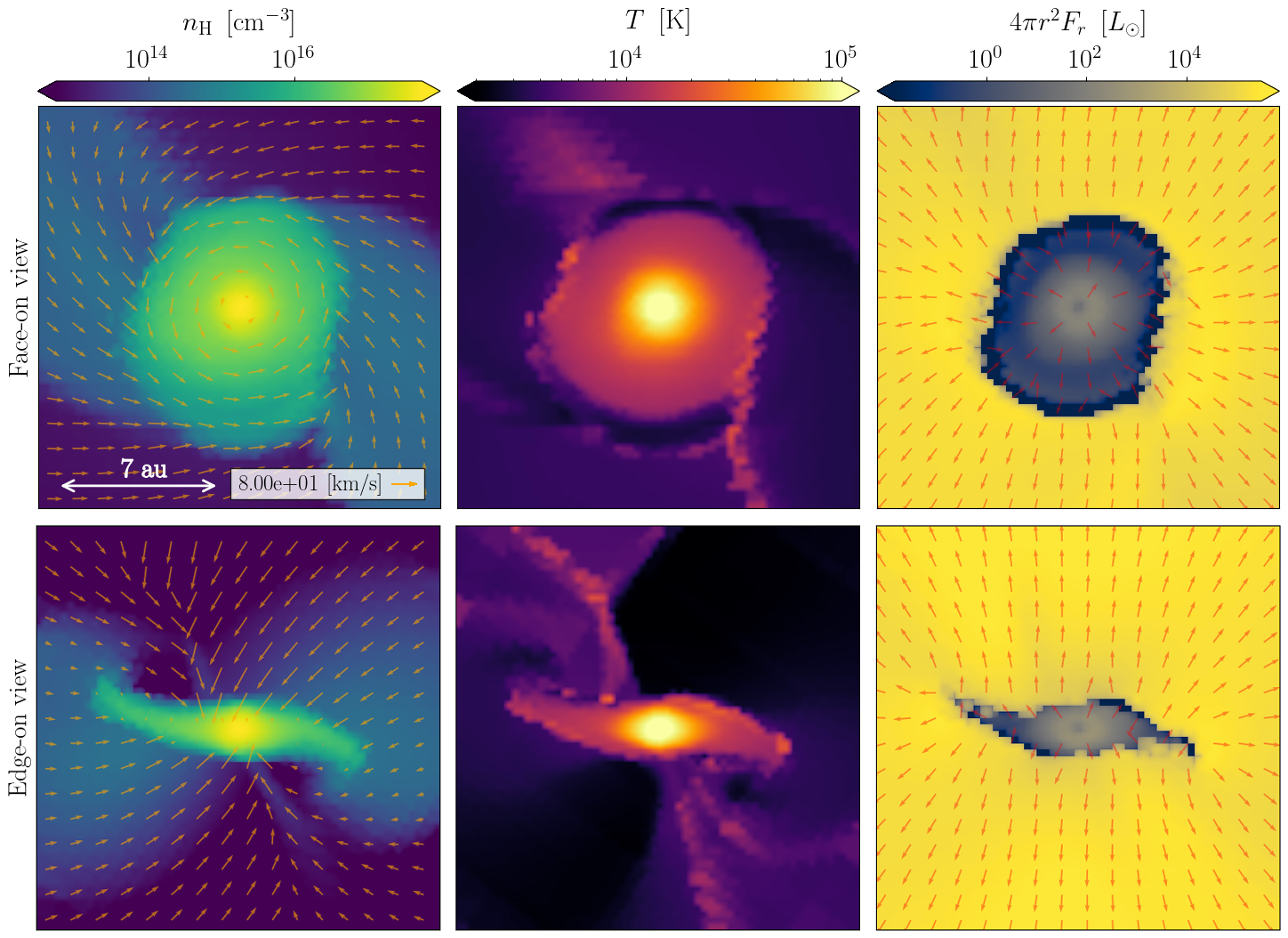}
    \caption{ 
        Snapshots at the end of the simulation, which follows the evolution of a protostar forming under a high accretion rate in the early Universe using the RHD scheme developed in this study.
        The upper and lower panels show face-on and edge-on views, respectively.
        The left, middle, and right columns display the sliced distributions of the density, temperature, and isotropic luminosity $L_\mr{iso}\equiv 4\pi r^2 F_r $, where $F_r$ is the local outward radiation flux. 
        The isotropic luminosity represents the flux intensity corrected for geometrical dilution. 
        In the density panels, orange arrows indicate velocity vectors, with lengths proportional to their magnitudes. 
        In the luminosity panels, red arrows represent the directions of the radiation flux.
    }
    \label{fig:ChonIC_2DSnap}
  \end{center}
\end{figure*}
\subsection{Setup}
To simulate the formation and evolution of a protostar, it is essential to accurately model the gas chemistry together with the associated heating and cooling processes.
For this purpose, we incorporate a primordial chemistry network into our code to follow the coupled thermal and chemical evolution of primordial gas \citep{Omukai_2001,Hosokawa_et_al_2016,Matsukoba_et_al_2019,Sugimura_et_al_2020}.
Specifically, the emission term $aT^4$ in $G_0$ and the opacities appearing in the governing equations are replaced with values computed from emissivities and opacities appropriate for primordial chemical compositions.
We adopt the same primordial chemistry network as in \citet{Kimura_et_al_2023}, which follows the abundances of H, H$_2$, H$^+$, H$^-$, H$_2^+$, and e$^-$.
\par
The approximation of representing the spectrum by a single-temperature Planck distribution introduced in Section~\ref{sec:spectral_reconstruction} is reasonable once a protostar has formed, and its radiation dominates the system. 
However, during the earlier stage when the entire cloud is still optically thin and collapsing, such an approximation is not sufficiently accurate because the radiation field has not reached equilibrium and deviates from the Planck distribution. 
Especially in the collapse phase of a primordial gas cloud, a Planck spectrum at a single effective temperature tends to overestimate the number of ionizing photons with energies above $13.6$ eV. 
Since the chemical and thermal evolution of primordial gas depends sensitively on the ionization degree, this overproduction of ionizing photons can strongly affect the thermal history. 
To avoid this, we divide the radiation into two frequency bins separated at $13.6$ eV. 
For each bin, we evolve both the radiation energy density and photon-number density and estimate the spectrum from the mean photon energy, as described in Section~\ref{sec:spectral_reconstruction}.
In computing the mean photon energy with Equation~\eqref{eq:get_Trad} and the opacities with Equations~\eqref{eq:kappa_0E}, \eqref{eq:kappa_0F}, \eqref{eq:kappa_0E_N} and \eqref{eq:kappa_0F_N}, we perform the relevant integrals over the frequency range corresponding to each bin.
With this treatment, the radiation naturally approaches a Planck-like spectrum after the protostar forms, while in the optically thin collapsing phase the resulting thermal and chemical evolution agrees with the expected physical behavior \citep{Omukai_2001}.
\par
As the initial condition, we adopt the results of a cosmological simulation performed by \citet{Chon_et_al_2016} with the $N$-body and smoothed particle hydrodynamics (SPH) code {\tt GADGET2} \citep{Springel_2005}.
This setup corresponds to the \textit{spherical cloud} model of \citet{Chon_et_al_2018}, who predict that rapid gas accretion at a rate of $\sim 1M_\odot~\mathrm{yr}^{-1}$ leads to the formation of a supermassive star exceeding $10^4~M_\odot$.
This formation pathway for supermassive stars is often referred to as the direct collapse scenario, in which the remnant massive black holes provide a promising explanation for the presence of supermassive black holes observed in the early Universe.
At the initial stage of our simulation, the gas has already become gravitationally unstable due to self-gravity, with a maximum number density of $10^4~{\rm cm}^{-3}$.
To generate the initial conditions for our grid-based code SFUMATO, 
we remap SPH data from {\tt GADGET2} onto Cartesian grids.
In the simulations of \citet{Chon_et_al_2018}, high-density regions were replaced with sink particles, and the internal structure of the protostar was not resolved.
Instead, protostellar evolution was incorporated as a subgrid model based on 1D stellar evolution calculations by \citet{Hosokawa_et_al_2012_SMS, Hosokawa_et_al_2013}.
In contrast, our simulations do not employ sink particles and resolve the protostellar interior directly with much higher spatial resolution.
The computational domain spans $\sim 100$~pc. 
The base grid consists of 64 cells in each direction, and adaptive mesh refinement is applied to ensure that the Jeans length is always resolved with at least 8 cells.
The finest spatial resolution achieved in our simulation is $4.7\times10^{-3}$~au.
\par
The extracted primordial gas cloud is irradiated by strong ultraviolet (UV) radiation from nearby star-forming galaxies, which suppresses H$_2$ formation and triggers gravitational collapse via hydrogen atomic cooling \citep{Chon_et_al_2018}. 
In our simulations, instead of explicitly including these galaxies within the computational domain, we mimic their effect by imposing a uniform background radiation field with an intensity of $1000~J_{21}$, where the conventional unit $J_{21}$ is defined as $10^{-21}~\mr{erg}~\mr{s}^{-1}~\mr{Hz}^{-1}~\mr{cm}^{-2}~\mr{sr}^{-1}$.
This radiation field induces a high photodissociation rate of H$_2$, thereby preventing molecular hydrogen formation \citep[e.g.][]{Sugimura_et_al_2014}.
\subsection{Results}
Once the simulation begins, the initial gas cloud undergoes gravitational collapse via hydrogen atomic cooling at a temperature of $\sim8000$~K, leading to the formation of a small optically thick protostar with a mass of $\sim0.01~M_\odot$. 
Subsequent accretion from the surrounding gas proceeds at a rate of $\sim 1~M_\odot~\mathrm{yr}^{-1}$. 
We follow the evolution for 10~yr after the protostellar formation. 
\par
In Figure~\ref{fig:ChonIC_2DSnap}, we present face-on and edge-on views of the protostar at the final snapshot of the simulation, showing the distributions of gas density, temperature, and luminosity.
The definition of luminosity is the same as that described in Section~\ref{subsec:Radiation_Escap}.
As seen in the figure, the gas can be divided into two distinct regions.
The outer region is characterized by low density and temperature, is optically thin, and exhibits high luminosity.
In contrast, the inner region, with a radius of $\sim 4$~au, is dense and hot, optically thick, and shows low luminosity.
As shown in the density panels, the gas in the outer region is infalling approximately in freefall with some angular momentum, while in the inner region it exhibits nearly pure rotational motion.
The mass of this dense, optically thick region is $\sim 15~M_\odot$.
From the right panels, the radiation emitted from this region has a luminosity of $\sim 3\times10^5~L_\odot$, which is consistent with that of a protostar of the same mass obtained in 1D stellar evolution calculations \citep{Hosokawa_et_al_2012_SMS}.
In the general context of star formation, the protostar and accretion disk are often regarded as distinct components.
In our simulation, however, only a single, rotating, optically thick object forms, and no clear boundary exists that separates the protostar from the surrounding accretion disk.
\par
Here, we discuss similarities and differences with previous studies. 
First, our findings are broadly consistent with those of \citet{Kimura_et_al_2023}, who started their simulations from an artificial Bonnor–Ebert sphere and likewise found the formation of a single, rotating, optically thick structure. 
In contrast, \citet{Luo_et_al_2018}, who adopted cosmological initial conditions, reported that the central protostar can be disrupted by gas and radiation pressure, accompanied by strong outflows, after which a new core forms at a different location.
Such features are not observed in our present calculation, and the origin of this discrepancy remains to be clarified in future work. 
We also note that \citet{Ahmad_et_al_2024}, in the context of present-day star formation, reported a smooth transition from the protostar to the inner disk, consistent with the structure found in our simulation.
\par
In summary, our application of the newly developed RHD scheme demonstrates its capability to resolve the formation and early evolution of a protostar under realistic conditions in the early Universe. 
Unlike previous approaches that relied on sink particles or 1D stellar evolution models, our simulation directly follows the protostar and its surrounding anisotropic accretion flow in 3D.
Extending such simulations to longer timescales will be crucial for understanding radiative and mechanical feedback, as well as the subsequent growth of the protostar, ultimately clarifying the nature of the final stellar object that forms.
\section{Discussion} \label{sec:Discussion}
\subsection{Local breakdown of energy conservation in optically thick system} \label{subsec:Local_break}
Although we introduce the non-RSLA component to achieve energy conservation in the optically thick system in Section~\ref{subsec:HybridRT},
energy conservation locally breaks down in optically thick systems if they contain regions where $\tau_\mathrm{cell} < 2 / (3f_c)$, since the RSLA component $\EofRSLA_\mathrm{rad}$ dominates in such regions.
This issue can become significant in extremely high-resolution simulations, where $\tau_\mathrm{cell}$ may fall below $2/(3 f_c)$ even within stellar interiors.
As discussed in Section~\ref{subsec:energy-nonconserve}, when energy is transferred from the gas to the radiation field, only a fraction $f_c$ of the transferred energy is actually received by the radiation.
In an extreme case where the gas energy is entirely converted into radiation, the RSLA reduces the total energy to only $f_c$ times its correct value.
At this time, because the radiation energy density scales as $T^4$, this reduction corresponds to a temperature deviation of $f_c^{1/4}$.
\par
However, in typical RHD star formation simulations, where the Jeans length is resolved by several tens of cells, $\tau_\mathrm{cell}$ remains sufficiently large throughout most of the stellar interior, and the regions with $\tau_\mathrm{cell} < 2 / (3f_c)$ are confined to narrow layers near the stellar surface.
As a result, the impact of the above effect on the global energy budget is expected to be small.
To illustrate this, consider representative photospheric conditions of the Sun:
\(T \simeq 5.8\times10^{3}\ {\rm K}\),
\(\rho \simeq 10^{-9}\ {\rm g\,cm^{-3}}\),
and a Rosseland--mean opacity
\(\kappa_\mr{0R} \simeq 1.5\times 10^{-2} \ {\rm cm^{2}\,g^{-1}}\).
The corresponding Jeans length is
$\lambda_\mr{J} \approx
c_\mr{s}\sqrt{\pi/(G\rho)}
\simeq 1.8\times10^{14}\ {\rm cm}$
while its optical depth is enormous,
$\tau_\mr{J} \sim \kappa_\mr{0R}\,\rho\,\lambda_\mr{J}
\simeq 2.7\times10^{3}$.
Thus, even if the Jeans scale were resolved by several tens of cells, each cell would still remain optically thick (\(\tau_{\rm cell} \gg 1\)) even near the photosphere.
Furthermore, as the density increases rapidly toward the stellar interior, $\tau_{\rm cell}$ also rises steeply and exceeds $2/(3f_c)$ near the surface.
\subsection{Radiation behavior in the transition region between the RSLA and non-RSLA components} \label{subsec:Transition_Region}
\begin{figure}[t]
  \begin{center}
    \includegraphics[width=\linewidth]{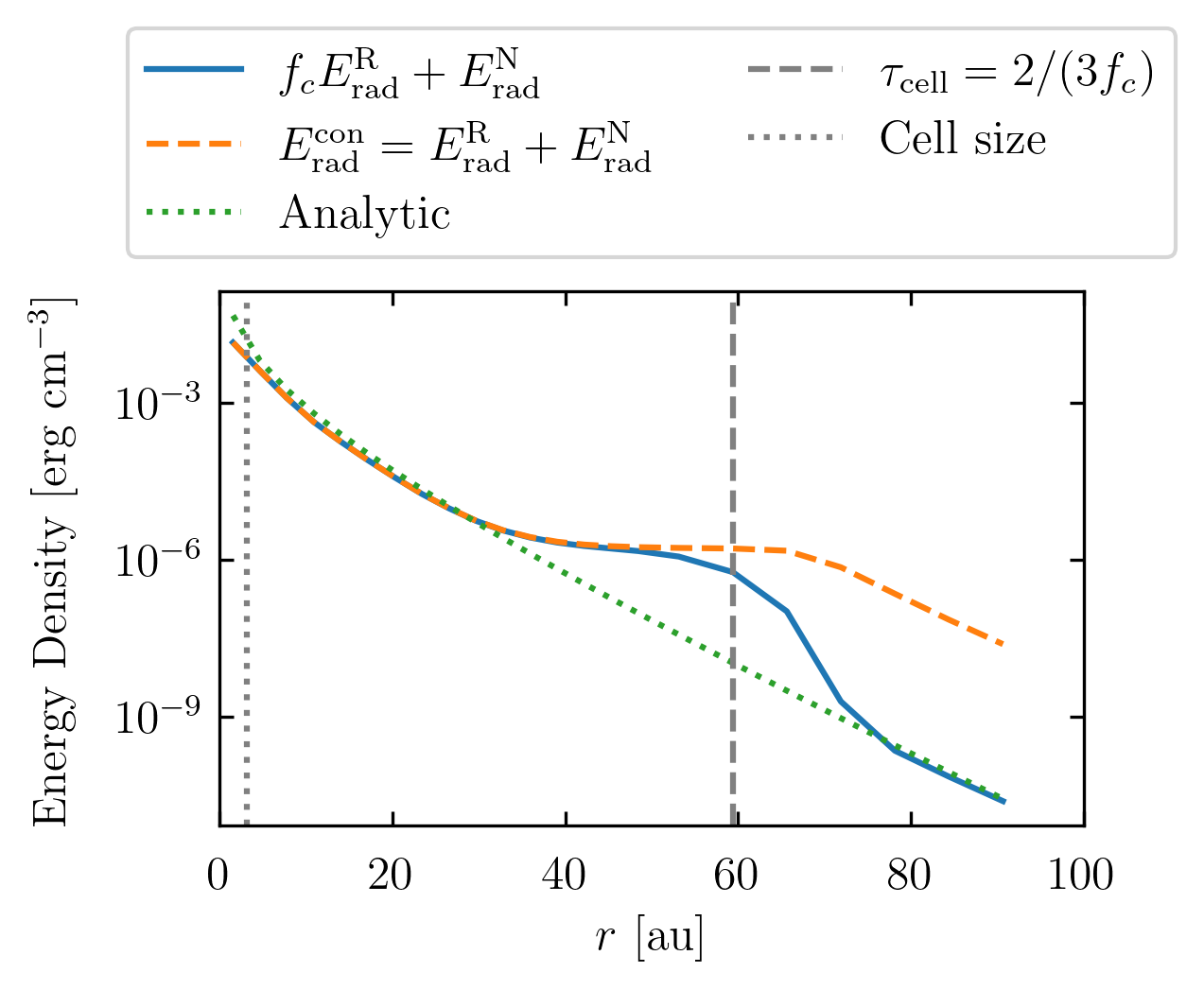}
    \caption{
        Transition test between the non-RSLA and RSLA components.
        The blue solid and orange-dashed lines represent the spherically averaged radial profile of physical radiation energy density and conserved radiation energy density in the steady state from our simulation.
        The green-dotted line denotes the analytic solution.
        The gray-dashed and -dotted lines indicate the radius at which $\tau_\mr{cell}=2/(3f_c)$ and the computational cell size ($3.13$~au).
        The simulation is performed in a cubic box of size $200$~au with a uniform grid of $64^3$.
    }
    \label{fig:Transition_Region}
  \end{center}
\end{figure}
In our new scheme, the radiation field is decomposed into the RSLA and non-RSLA components according to Equation~\eqref{eq:distribute_RN} as mentioned in Section~\ref{subsec:HybridRT}.
With this decomposition, we can solve RT using the RSLA component in the optically thin limit and the non-RSLA component in the optically thick limit.
However, in the intermediate regime, the radiation field is constructed by interpolating between the optically thin and optically thick limits.
In this section, we examine how the radiation field behaves in this transition region.
We perform the calculation in which the radiation is injected at the center of the computational box and transported outward, using the same setup as in Section~\ref{subsec:Diffusion}, except for the spatial resolution and the gas density distribution.
For the resolution, the computational box width is $200$~au, and we resolve the box with $64^3$ cells, corresponding to a cell size of $3.13~\mathrm{au}$.
For the gas density distribution, we adopt an exponential gas density profile,
\begin{equation}
\rho(r) = A \exp(-r/B)~\mathrm{g~cm^{-3}},
\end{equation}
where $r$ is the radial distance measured in centimeters, $A = 2.7\times10^{18}$ and $B =9\times 10^{13}$~cm.
As a result, the optical depth per cell $\tau_{\mathrm{cell}}$ is $10^{7}$ at the center of the computational box and decreases to unity at the outer boundary.
We set $f_c = 10^{3}$, so that the transition between the dominance of the non-RSLA and RSLA components occurs at $\tau_{\mathrm{cell}} = 2/(3 f_c) \simeq 667$, which corresponds to a radius of $60~\mathrm{au}$.
In this case, assuming a steady state, we can obtain an analytic solution in the same manner as in Section~\ref{subsec:Diffusion},
\begin{eqnarray}
E_{\mathrm{rad}} = \frac{3\sigma_0^\nu A L}{4\pi c}
\left[
\frac{e^{-r/B}}{r}
+ \frac{1}{B}\,\mathrm{Ei}\!\left(-\frac{r}{B}\right)
\right], \label{eq:ana_exp}
\end{eqnarray}
where $\mathrm{Ei}$ denotes the exponential integral.
In the calculation, we impose Dirichlet boundary conditions using the value given by Equation~\eqref{eq:ana_exp}.
\par
Figure~\ref{fig:Transition_Region} shows the spherically averaged radial profiles of the physical radiation energy density,
$f_c \EofRSLA_\mathrm{rad} + \EofNonRSLA_\mr{rad}$ (the blue solid line), and the conserved radiation energy density,
$E^\mr{con}_\mr{rad}=\EofRSLA_\mr{rad}+\EofNonRSLA_\mr{rad}$
(the orange-dashed line), in the steady state.
For comparison, the analytic solution is shown (the green-dotted line).
We also indicate the radius at which $\tau_{\mathrm{cell}} = 2/(3 f_c)$ and the cell size by the gray-dashed and -dotted lines, respectively.
Near the center ($r < 3.13~\mathrm{au}$), a small discrepancy between the simulation result and the analytic solution is present, which arises from the finite size of the injection cell, as discussed in Section~\ref{subsec:Diffusion}.
\par
In Figure~\ref{fig:Transition_Region}, we show the conserved radiation energy density $E^{\mathrm{con}}_{\mathrm{rad}}$ by the orange-dashed line.
In the inner region ($r < 30~\mathrm{au}$), where the non-RSLA component is dominant and $E^{\mathrm{con}}_{\mathrm{rad}} \simeq \EofNonRSLA_\mr{rad}$, $E^{\mathrm{con}}_{\mathrm{rad}}$ follows the analytic solution.
In contrast, in the outer region ($r > 80~\mathrm{au}$), where the RSLA component dominates and $E^{\mathrm{con}}_{\mathrm{rad}} \simeq \EofRSLA_\mr{rad}$, $E^{\mathrm{con}}_{\mathrm{rad}}$ follows an analytic solution elevated by a factor of $1/f_c$.
This enhancement arises because, when the RSLA is effective, the radiation propagation speed is suppressed, leading to a larger accumulation of radiation energy in a steady state.
The orange-dashed line in Figure~\ref{fig:Transition_Region} connects these inner and outer solutions.
\par
The physical radiation energy density, $f_c \EofRSLA_\mathrm{rad} + \EofNonRSLA_\mr{rad}$ (the blue solid line in Figure~\ref{fig:Transition_Region}), agrees well with the analytical solution in both the inner and outer regions.
In the outer region where RSLA-component $\EofRSLA_\mr{rad}$ is dominant, the physical radiation energy density is approximated as $f_c E^{\mathrm{con}}_{\mathrm{rad}}$ and consistent with the analytic solution.
Then, the physical radiation energy density gradually approaches $E^{\mathrm{con}}_{\mathrm{rad}}$ toward smaller radii since the fraction of non-RSLA component $\EofNonRSLA_\mr{rad}$ becomes larger.
In the intermediate region, however, $E^{\mathrm{con}}_{\mathrm{rad}}$ has not yet fully converged to the analytic solution, which results in a deviation of the physical radiation energy density from the analytic solution.
At sufficiently small radii, $E^{\mathrm{con}}_{\mathrm{rad}}$ converges to the analytic solution, and consequently, the physical radiation energy density also agrees with it.
\par
As demonstrated in this test, our scheme successfully reproduces the physically correct state in both the inner and outer regions. 
However, a discrepancy from the physically correct state remains in the transition region between the non-RSLA and RSLA components.
For the present application focusing on protostellar evolution, this transition layer is confined to an extremely thin region near the protostellar surface, and this discrepancy is therefore not expected to have a significant impact on our results. 
In fact, in Section~\ref{subsec:Radiation_Escap}, we demonstrate that our scheme can accurately follow radiative transport across regions where the optical depth changes abruptly. 
Moreover, this discrepancy can be mitigated by choosing the reduction factor $f_c$ closer to unity.
\subsection{Choice of reduction factor $f_c$}
In our scheme, $f_c$ is a free parameter.
The choice of $f_c$ is typically guided by the requirement that the ordering between 
the gas acoustic signal speed $v_{\mathrm{sig}} = |\boldsymbol{v}| + c_{\mathrm{eff}}$ 
(where $c_{\mathrm{eff}}$ is the effective sound speed including radiation pressure) 
and the effective radiation propagation speed $\tilde{c}/\max(\tau,1)$ should not be reversed relative to the ordering in the non-RSLA system \citep{Skinner_and_Ostriker_2013}. 
For the protostellar evolution problems considered in this work, we choose $f_c$ such that the correct ordering is maintained in the optically thin regions outside the protostar, while allowing a reversal of the ordering in the dense interior of the protostar.
Generally, when a simulation is designed to resolve both the exterior and interior of a protostar, it is not possible to satisfy this ordering criterion throughout the entire computational domain.
This is because the effective radiation propagation speed is larger in the optically thin exterior, whereas the gas effective sound speed is larger in the dense interior.
As a result, for any choice of $f_c$, the RSLA inevitably introduces an artificial reversal of these characteristic speeds somewhere between the exterior and interior.
However, because the optical depth rises extremely steeply across the protostellar surface, our choice ensures that any such artificial reversal is confined to narrow layers near the surface, and we therefore expect it to have only a minor influence on the global protostellar evolution captured in our simulations.
\section{Summary} \label{sec:Summary}
In this work, we have developed a novel RHD scheme for simulating the formation and evolution of protostars together with their surrounding accretion flows.
Our scheme is based on the explicit M1 closure method coupled with the RSLA. This formulation efficiently captures the anisotropy of radiation fields while maintaining high computational performance on massively parallel architectures, making it well suited for long-term, large-scale simulations of star formation.
\par
The primary motivation for developing the new scheme is to overcome three major limitations inherent to the explicit M1 closure method with the RSLA.
First, conventional RSLA treatments inevitably violate total energy conservation in optically thick, radiation-dominated regions, hindering accurate modeling of stellar evolution (Section~\ref{subsec:energy-nonconserve}).
Second, the standard M1 scheme suffers from excessive numerical diffusion in highly optically thick stellar interiors or fails to allow radiation to escape properly across steep optical-depth gradients, such as at protostellar surfaces (Section~\ref{subsec:RT_in_transition}).
Third, most existing RHD simulations employ overly simplified spectral assumptions, which prevent them from following the correct temporal and spatial evolution of the radiation spectrum and lead to inaccurate estimates of protostellar radiative feedback (Section~\ref{subsec:overview_spectrum}).
\par
In our new scheme, we overcome the above three limitations as follows.
To address the first issue, we introduce a hybrid RT approach where both RSLA and non-RSLA radiation components are evolved simultaneously (Section~\ref{subsec:HybridRT}). The non-RSLA component dominates in optically thick regions and restores total energy conservation.
For the second issue, we improve the treatment of RT by redistributing the radiation energy among the streaming, trapped, and non-RSLA components using both local and neighboring optical-depth information (Section~\ref{subsec:RadPartition}).
This method allows radiation to diffuse correctly within stellar interiors and to escape properly across steep optical-depth gradients at protostellar surfaces.
To resolve the third issue, we evolve photon-number densities in addition to the radiation energy densities, which enables us to reconstruct the local spectrum in each cell consistently with the evolving protostar (Section~\ref{sec:spectral_reconstruction}).
This, in turn, allows for a more accurate evaluation of radiative feedback.
\par
We have implemented this new scheme in SFUMATO, a self-gravitating magnetohydrodynamics code with adaptive mesh refinement, which has been widely applied to studies of star formation.
The implementation is based on an operator-splitting procedure, in which hydrodynamics, self-gravity, radiation transport, emission and absorption, and momentum exchange are solved step by step in a stable manner (Section~\ref{sec:comp_proc}).
\par
We have validated the method through a series of numerical tests (Section~\ref{sec:tests}), including energy conservation, RT in optically thick regimes, radiative diffusion across steep optical-depth gradients, and reconstruction of evolving radiation spectra.
These tests demonstrate that the scheme performs robustly and accurately under conditions relevant to protostellar formation and evolution.
\par
As an application of the new scheme, we have investigated cosmological protostellar evolution in the early Universe under the so-called direct collapse scenario, where gas accretion rates are extremely high, within a full cosmological context (Section~\ref{sec:ProsForm}).
Our simulations resolve the protostellar interior together with the large-scale anisotropic accretion flow, and demonstrate that the protostar grows while maintaining a smooth connection to its surrounding disk.
Such a continuous star–disk interface implies that entropy is not generated abruptly at the stellar surface but is supplied gradually through dissipative processes within the disk, which may affect its radial inflation.
The overall evolutionary path we obtained is in close agreement with that reported in our previous study \citep{Kimura_et_al_2023}, which started from an artificial Bonnor–Ebert sphere.
These features cannot be obtained in conventional 1D evolution models, and this application demonstrates that our method can capture protostellar evolution in realistic environments.
\par
The versatility of our scheme allows it to be applied to a wide range of astrophysical problems. In the context of protostellar accretion, magnetic fields may play an important role in the region extending from the inner accretion disk to the stellar surface \citep[e.g.,][]{Mayer_et_al_2025_arXiv,Takasao_et_al_2025}. To explore such effects, our framework can be extended toward full radiation magnetohydrodynamics simulations that incorporate magnetic fields self-consistently. The same framework is also relevant to other problems in stellar physics (\citet{Jiang2023}; and references therein), such as mass loss from massive stars, including Wolf–Rayet winds \citep[][]{Moens_et_al_2022}, where the flow direction is opposite to that in protostellar accretion. On even larger spatial scales, our method can be applied to galactic or circumnuclear environments where radiation pressure and gravity interact to produce complex gas dynamics involving both inflows and outflows \citep[][and references therein]{Thompson_and_Heckman_2024}. Such systems, in which radiation feedback regulates accretion and can even trigger star formation in dense regions, provide another natural target for our scheme. Together, these examples demonstrate the broad applicability of our method and its potential to advance our understanding of RHD processes across diverse astrophysical environments.

\begin{acknowledgments}
The authors would like to thank Tomoaki Matsumoto, Kengo Tomida, and Chong-Chong He for fruitful discussions and valuable comments. 
We also thank Sunmyon Chon for providing the simulation data used in this work and for helpful discussions.
T.H. wishes to express his cordial gratitude to Prof. Takahiro Tanaka for his continuous interest and encouragement.
H.F. wishes to express his heartfelt gratitude to his most revered mentor, Prof. Masayuki Umemura, the Pater Patriae of the Theoretical Astrophysics Group at the University of Tsukuba, as well as to Prof. Ken Ohsuga, for their constant interest, valuable advice, and encouragement.
The numerical simulations were carried out on the XC50 (Aterui II) and XD2000 (Aterui III) at the Center for Computational Astrophysics (CfCA), National Astronomical Observatory of Japan, through the courtesy of Prof. Eiichiro Kokubo, and on the Yukawa-21 at the Yukawa Institute for Theoretical Physics, Kyoto University.
This research was supported in part by Grants-in-Aid for Scientific Research (KK: 24KJ0015, KS: 22KK0043, 24H00002, TH: 19KK0353, 22H00149, HF: 23K13139, KO: 22H00149) from the Japan Society for the Promotion of Science, and by the Astrobiology Center Project research (HF: AB0511). T.H. also appreciates the financial support from ISHIZUE 2024 provided by Kyoto University and the Kyoto University Foundation.
\end{acknowledgments}

\section*{Data Availability}
The input files used for the numerical tests and simulation data are available from the corresponding author upon reasonable request.

\bibliography{bib}{}
\bibliographystyle{aasjournal}



\appendix
\section{Lorentz transformations} \label{sec:app_LT}
We introduce the Lorentz transformations of the radiation quantities in the limit of $\beta\ll1$. The radiation energy density $E_\mathrm{rad}$, radiation flux $\bm{F}$, and radiation-pressure tensor $\ten{P}_\mathrm{rad}$ in the lab frame are related to their comoving-frame counterparts $E_{\mathrm{rad},0}$, $\bm{F}_0$, and $\ten{P}_{\mathrm{rad},0}$ as
\begin{eqnarray}
    && E_\mr{rad} = E_\mr{rad,0} + 2 \bra{\f{\bm{v}}{c}}\cdot\f{\bm{F}_0}{c} , \label{eq_Lorentz:E} \\
    && \f{\bm{F}}{c} = \f{\bm{F}_0}{c} + \f{\bm{v}}{c}\cdot(E_\mr{rad,0}\ten{I}+\ten{P}_\mr{rad,0}) , \label{eq_Lorentz:F} \\
    && \ten{P}_\mr{rad} = \ten{P}_\mr{rad,0} + \bra{\f{\bm{v}}{c}\otimes \f{\bm{F}_0}{c}+\f{\bm{F}_0}{c}\otimes\f{\bm{v}}{c}} , \label{eq_Lorentz:P}
\end{eqnarray}
which are obtained by retaining terms up to first order in $\beta$ in the Lorentz transformations (see Equations 91.10 and 91.11 in \citeauthor{Mihalas_and_Mihalas_1984}~\citeyear{Mihalas_and_Mihalas_1984}).
In Equation~\eqref{eq_Lorentz:F}, the second term on the right-hand side represents the advection of radiation enthalpy by the moving gas.
\par
Here, we estimate the relative magnitudes of the lab frame and comoving-frame quantities in the limit of $\beta\ll 1$.
Since $|\bm{F}_0| \leq c E_{\mathrm{rad},0}$, it follows from Equations \eqref{eq_Lorentz:E} and \eqref{eq_Lorentz:P} that the differences between the lab frame and comoving-frame quantities of energy density and radiation-pressure tensor are at most of the order $\mathcal{O}(\beta)$. 
Therefore, to the leading order, these differences can be neglected, and we can assume
\begin{eqnarray}
    E_\mr{rad} \simeq E_\mr{rad,0} , \hspace{5mm} \ten{P}_\mr{rad} \simeq \ten{P}_\mr{rad,0} . \label{eq_app:EP_identical}
\end{eqnarray}
On the other hand, even with $\beta\ll1$, the difference between $\bm{F}$ and $\bm{F}_{0}$ cannot be negligible in the dynamic diffusion limit where $\tau\gg1$ and $\tau\beta\gg1$. When $\tau\gg1$, the comoving flux $\bm{F}_0$ can be described by the diffusion approximation as
\begin{eqnarray}
    \bm{F}_0 = -\f{c}{3\chi_\mr{0R}\rho}\nabla E_\mr{rad,0}\sim\mathcal{O}\bra{\f{c}{\tau} E_\mr{rad,0}}.
\end{eqnarray}
Accordingly, the order of magnitude of $(\bm{v}/c)\cdot(E_\mr{rad,0}\ten{I}+\ten{P}_\mr{rad,0})$ in Equation~\eqref{eq_Lorentz:F} differs from that of $\bm{F}_0/c$ by a factor of $\tau\beta$, and thus becomes dominant in the dynamic diffusion regime.
\par
Furthermore, since the second term on the right-hand side of Equation~\eqref{eq_Lorentz:F} contributes only in the dynamic diffusion regime, we can adopt the relation $\ten{P}_\mathrm{rad,0} = E_\mathrm{rad,0} \ten{I} / 3$ for optically thick gas, leading to 
\begin{eqnarray}
&& \f{\bm{F}}{c} = \f{\bm{F}_0}{c} + \f{4}{3}\bm{\beta}E_\mr{rad,0} . \label{eq_Lorentz:F2}
\end{eqnarray}
For simplicity, in this paper, we often use this equation instead of Equation~\eqref{eq_Lorentz:F}.
\section{Mixed frame RHD Equations at Leading Order} \label{sec:app_dom}
In this appendix, we derive the set of equations that should be solved when the gas velocity is much smaller than the speed of light, $\beta\ll1$. 
A key point to note is that we aim to perform the simulation resolving regions inside protostars, where the density is high, and the medium is extremely optically thick ($\tau \gg 1$). 
In this case, terms proportional to $\beta$ are not necessarily negligible since the product $\tau \beta$ can be very large, which is the so-called dynamic diffusion regime.
Therefore, we must take special care in the derivation of the RHD equations.
A detailed analysis of the magnitude of the terms appearing in the RHD equations under the assumption of $\beta\ll1$ has been carried out by \citet{Krumholz_et_al_2007_Algorithms} and \citet{Skinner_and_Ostriker_2013}.
They examine the relative magnitudes of each term in the equations across various regimes, including the optically thin limit ($\tau\ll1$), the static diffusion limit ($\tau\gg1$ and $\tau\beta\ll1$), and the dynamic diffusion limit ($\tau\gg1$ and $\tau\beta\gg1$).
In this paper, based on their analysis, we derive the RHD equations by retaining only the leading-order terms.
\par
First, by retaining the terms up to the second order in $\beta$ in Equations (54b) and (54d) of \citet{Mihalas_and_Auer_2001}, we obtain the radiation four-force $(G^0,\bm{G})$ density as
\begin{eqnarray}
    && G^0 = \rho ( \kappa_{0E} E_\mr{rad}
    - \kappa_\mr{0P} a T^4) + \rho (\chi_{0F} - 2 \kappa_{0E}) \f{\bm{\beta}\cdot\bm{F}}{c} \nonumber \\
    && \hspace{1cm} - \rho(\chi_{0F}-\kappa_{0E}) \bra{\beta^2 E_\mr{rad}+(\bm{\beta}\otimes\bm{\beta}):\ten{P}_\mr{rad}} + \f{1}{2}\beta^2\rho(\kappa_{0E}E_\mr{rad}-\kappa_\mr{0P}a T^4) , \label{eq:G0_first} \\
    && \bm{G} = \rho\chi_{0F}\f{\bm{F}}{c} - \rho\chi_{0F} \bm{\beta} \cdot (E_\mr{rad}\ten{I}+\ten{P}_\mr{rad}) +\bm{\beta}\rho(\kappa_{0E}E_\mr{rad}-\kappa_\mr{0P}a T^4) \nonumber \\
    && \hspace{1cm}  + \f{1}{2}\beta^2\rho\chi_{0F}\f{\bm{F}}{c} + 2\rho(\chi_{0F}-\kappa_{0E})\f{(\bm{\beta}\cdot\bm{F})\bm{\beta}}{c}. \label{eq:G_first}
\end{eqnarray}
Considering the optically thick regime and replacing $\kappa_{0E}$ and $\chi_{0F}$ (Equations~\ref{eq:kappa_0E} and \ref{eq:kappa_0F}) with $\kappa_{0\mathrm{P}}$ and $\chi_{0\mathrm{R}}$ (Equations~\ref{eq:kappa_0P} and \ref{eq:Rosseland}) make these equations equivalent to Equations (24) and (25) in \citet{Krumholz_et_al_2007_Algorithms}.
\par
\begin{table}[t]
    \caption{
         Scalings of the terms in the radiation four-force density $(G^0, \bm{G})$ in different limiting regimes. 
         All scalings are normalized by $E_{\mathrm{rad},0}/l$, where $l$ is the characteristic length scale of the system. 
         The dominant terms in each regime are indicated by an asterisk.
    }
    \label{tab:scalings}
    \centering
    \begin{tabular}{lcccc}
        \hline \hline
        Term & $G^0$ or $\bm{G}$ & Streaming & Static diffusion & Dynamic diffusion \\ \hline \hline
        $\rho ( \kappa_{0E} E_\mr{rad} - \kappa_\mr{0P} a T^4)$ & $G^0$ & $\tau^*$ & $1/\tau^* $ & $\beta^2\tau^*$ \\
        $\rho (\chi_{0F} - 2 \kappa_{0E}) (\bm{\beta}\cdot\bm{F}/c)$ & $G^0$ & $\beta\tau$ & $\beta$ & $\beta^2\tau^*$ \\
        $- \rho(\chi_{0F}-\kappa_{0E}) \bra{\beta^2 E_\mr{rad}+\bm{\beta}\bm{\beta}:\ten{P}_\mr{rad}}$ & $G^0$ & $\beta^2\tau$ & $\beta^2\tau$ & $\beta^2\tau^*$ \\
        $(1/2)\beta^2\rho(\kappa_{0E}E_\mr{rad}-\kappa_\mr{0P}a T^4)$ & $G^0$ & $\beta^2\tau$ & $\beta^2/\tau$ & $\beta^4\tau$
        \\ \hline
        $\rho\chi_{0F}(\bm{F}/c)$ & $\bm{G}$ & $\tau^*$ & $1^*$ & $\beta\tau^*$ \\
        $- \rho\chi_{0F} \bm{\beta} \cdot (E_\mr{rad}\ten{I}+\ten{P}_\mr{rad})$ & $\bm{G}$ & $\beta\tau$ & $\beta\tau$ & $\beta\tau^*$ \\
        $\bm{\beta}\rho(\kappa_{0E}E_\mr{rad}-\kappa_\mr{0P}a T^4)$ & $\bm{G}$ & $\beta\tau$ & $\beta/\tau$ & $\beta^3\tau$ \\ 
        $(1/2) \beta^2 \rho\chi_{0F}(\bm{F}/c)$ & $\bm{G}$ & $\beta^2\tau$ & $\beta^2$ & $\beta^3\tau$ \\
        $2\rho(\chi_{0F}-\kappa_{0E}) ((\bm{\beta}\cdot\bm{F})\bm{\beta}/c)$ & $\bm{G}$ & $\beta^2\tau$ & $\beta^2$ & $\beta^3\tau$ \\ \hline
    \end{tabular}
\end{table}
In Table~\ref{tab:scalings}, we present the scaling of each term in Equations~\eqref{eq:G0_first} and \eqref{eq:G_first}.
Our scaling analysis yields results identical to those in Table~1 of \citet{Krumholz_et_al_2007_Algorithms}, except that we adopt mean opacities for general spectra instead of those for thermal spectra considered in \citet{Krumholz_et_al_2007_Algorithms}.
For $G^0$ in the dynamic diffusion regime, the first, second, and third terms on the right-hand side of Equation~\eqref{eq:G0_first} can all become dominant.
However, it is not necessary to solve the second and third terms if the goal is to obtain the radiation energy density $E_{\mathrm{rad}}$ to leading order.
As shown in the first row of Table~\ref{tab:scalings}, the quantity $\rho(\kappa_{0E}E_{\mathrm{rad}} - \kappa_{0\mathrm{P}}aT^4)(l/E_\mr{rad,0})$ scales as $\mathcal{O}(\beta^2 \tau)$.
Assuming $\tau \gg 1$ and $\kappa_{0E} = \kappa_{0\mathrm{P}}$, this implies that $E_{\mathrm{rad}}$ deviates from $aT^4$ only by an amount of order $\mathcal{O}(\beta^2 E_\mr{rad,0})$.
While resolving this small deviation requires inclusion of the second and third terms, such accuracy is unnecessary if one is concerned only with the leading-order behavior.
Therefore, the second and third terms can be neglected in that case.
On the other hand, for the momentum source term $\bm{G}$ given by Equation~\eqref{eq:G_first}, the dominant contributions arise from the first and second terms on the right-hand side.
Thus, we need to retain both of these terms.
\par
Based on the above discussion, if one is interested only in the leading-order solution, it is sufficient to solve the RHD equations with the radiation four-force density $(G^0, \bm{G})$ given by
\begin{eqnarray}
    && G^0 = \rho ( \kappa_{0E} E_\mr{rad} - \kappa_\mr{0P} a T^4) , \label{eq:G0_leading}\\
    && \bm{G} = \rho\chi_{0F}\f{\bm{F}}{c} - \rho\chi_{0F} \bm{\beta} \cdot (E_\mr{rad}\ten{I}+\ten{P}_\mr{rad}) . \label{eq:G_leading}
\end{eqnarray}
Note that the right-hand side in Equation~\eqref{eq:G_leading} can be expressed with comoving-frame flux $\bm{F}_0$ as $\rho\chi_{0F}\bm{F}_0/c$ from Equation~\eqref{eq_Lorentz:F}.
Furthermore, since the second term on the right-hand side of Equation~\eqref{eq:G_leading} contributes only in the dynamic diffusion regime (i.e., when $\tau\gg1$ and $\tau\beta\gg1$), we can adopt $\ten{P}_\mr{rad}=E_\mr{rad}\ten{I}/3$ for optically thick gas, leading to
\begin{eqnarray}
    \bm{G} =  \rho\chi_{0F}\f{\bm{F}}{c} - \rho\chi_{0F} \f{4\bm{\beta}}{3} E_\mr{rad}. \label{eq:G_final}
\end{eqnarray}
Equations~\eqref{eq:G0_leading} and \eqref{eq:G_final} are identical to Equations~\eqref{eq_basic:G0} and \eqref{eq_basic:G}.

\section{Equations for Photon-Number Density} \label{sec:NumberDensity}
Here, we present the equations governing the photon-number density. 
By analogy with Equations~\eqref{eq:tildeE}--\eqref{eq:tildeG} for the radiation energy density, the photon-number density evolves according to
\begin{eqnarray}
   && \f{\p N^\mr{con}_\mr{rad}}{\p t} + \nabla\cdot\left\{\bm{J}_\mr{S}-\f{c}{3\hat{\chi}^\prime\rho}\nabla \NofNonRSLA_\mr{rad} + \f{4\bm{v}}{3}(\NofRSLA_\mr{rad,T}+\NofNonRSLA_\mr{rad})\right\} = -cG^{\prime 0} , \label{eq:Nrad} \\
   && \f{\p \bm{J}_\mr{S}}{\p t} + \tilde{c}^2\nabla\cdot\tenQofRSLA_\mr{rad,S} = -c\tilde{c} \bmGofRSLAprime_\mr{S} , \\
   && G^{\prime 0} = \rho \left\{  \kappa_{0E}^\prime (f_c \NofRSLA_\mr{rad}+\NofNonRSLA_\mr{rad}) - \f{30\zeta(3)}{\pi^4}\f{\kappa_\mr{0P}^\prime a T^4}{k_\mr{B} T} \right\}, \label{eq:Gprime0} \\
   && \bmGofRSLAprime_\mr{S} = \rho \chi_{0F}^\prime \f{\bm{J}_\mr{S}}{c} - \rho \chi_{0F}^\prime \f{4\bm{v}}{3c} f_c \NofRSLA_\mr{rad,S} .  \label{eq:bmGprime}
\end{eqnarray}
Here, the photon-number density variables, $N^\mr{con}_\mr{rad},\NofRSLA_\mr{rad}$, $\NofRSLA_\mr{rad,S}$, $\NofRSLA_\mr{rad,T}$, and $\NofNonRSLA_\mr{rad}$, correspond to the radiation energy-density variables, $E^\mr{con}_\mr{rad},\EofRSLA_\mr{rad}$, $\EofRSLA_\mr{rad,S}$, $\EofRSLA_\mr{rad,T}$, and $\EofNonRSLA_\mr{rad}$. 
Similarly, $\bm{J}_\mr{S}$ and $\tenQofRSLA_\mr{rad,S}$ correspond to $\bm{F}_\mr{S}$ and $\PofRSLA_\mr{rad,S}$, respectively.
For the second term in Equation~\eqref{eq:Gprime0}, we use the frequency-integrated photon-number emissivity, given by
\begin{eqnarray}
    && 4\pi \int_0^\infty d\nu \, \frac{ \rho \kappa^\nu_{0} B_\nu(T) }{ h \nu }
    = \frac{30 \, \zeta(3)}{ \pi^4 } \, \frac{ \rho \kappa_\mathrm{0P}^\prime c a T^4 }{ k_\mathrm{B} T } , \\
    && \kappa_\mathrm{0P}^\prime = \frac{ \int_0^\infty d\nu \, \kappa^\nu_{0} B_\nu(T)/h\nu }{ \int_0^\infty d\nu \, B_\nu(T)/h\nu } ,
\end{eqnarray}
where $\kappa_\mathrm{0P}^\prime$ is the Planck mean opacity weighted by the photon-number spectrum.
This expression is derived from Kirchhoff's law, which states that the emission coefficient is equal to the product of the absorption coefficient and the blackbody intensity.
Furthermore, we define frequency-integrated opacities for the photon-number density as follows:
\begin{eqnarray}
    && \kappa_{0E}^\prime = \f{\int_0^\infty d\nu \h \kappa^\nu_{0} E^\nu_{\mr{rad},0}/ h\nu}{\int_0^\infty d\nu \h E^\nu_{\mr{rad},0}/ h\nu} , \label{eq:kappa_0E_N} \\
    && \chi_{0F}^\prime = \f{\int_0^\infty d\nu \h (\kappa^\nu_{0}+\sigma^\nu_{0}) F^\nu_{0}/ h\nu}{\int_0^\infty d\nu \h F^\nu_{0}/ h\nu} . \label{eq:kappa_0F_N}
\end{eqnarray}
We estimate $\chi^\prime_{0F}$ in the same manner as $\chi_{0F}$ in Section~\ref{sec:spectral_reconstruction}, obtaining
\begin{eqnarray}
    && \chi_{0F}^\prime = \f{3(1-\xi_\chi)}{2}\chi_\mr{0R}^\prime + \f{3\xi_\chi-1}{2}\chi_{0E}^\prime , \label{eq:kappa_0F_N_exp} \\
    && \chi_{0E}^\prime = \f{\int_0^\infty d\nu \h (\kappa^\nu_{0}+\sigma^\nu_{0}) E^\nu_{\mr{rad},0}/ h\nu}{\int_0^\infty d\nu \h E^\nu_{\mr{rad},0}/ h\nu} , \\
    && \f{1}{\chi_\mr{0R}^\prime} = \f{\int_0^\infty d\nu \h (\kappa^\nu_{0}+\sigma^\nu_{0}) (\partial B^\nu(T)/\partial T)/h\nu}{\int_0^\infty d\nu \h (\partial B^\nu(T)/\partial T)/h\nu} ,
\end{eqnarray}
where $\xi_\chi$ is given by Equation~\eqref{eq:xi_chi}.
We obtain $\tenQofRSLA_\mr{rad,S}$ by imposing the M1 closure relation,
\begin{eqnarray}
    && \tenQofRSLA_\mr{rad,S} = \ten{D}^\prime_\mr{S}\NofRSLA_\mr{rad,S} , \nonumber \\
    && \ten{D}^\prime_\mr{S} = \f{1-\chirad^\prime_\mr{S}}{2}\ten{I} + \f{3\chirad^\prime_\mr{S}-1}{2}\bm{n}^\prime_\mr{S}\otimes\bm{n}^\prime_\mr{S}, \\
    && \bm{n}^\prime_\mr{S}=\f{\bm{J}_\mr{S}}{|\bm{J}_\mr{S}|}, \h
    \chirad^\prime_\mr{S}=\f{3+4(|\bm{J}_\mr{S}|/(\tilde{c}\NofRSLA_\mr{rad,S}))^2}{5+2\sqrt{4-3(|\bm{J}_\mr{S}|/(\tilde{c}\NofRSLA_\mr{rad,S}))^2}}
    , \nonumber 
\end{eqnarray}
which is equivalent to Equation~\eqref{eq:tilde_D_S}.
\section{Impact of a Nonconservative Self-gravity Scheme on Protostellar Evolution} \label{subsec:Com_Self_Gravity}
\begin{figure}[t]
  \begin{center}
    \includegraphics[width=\linewidth]{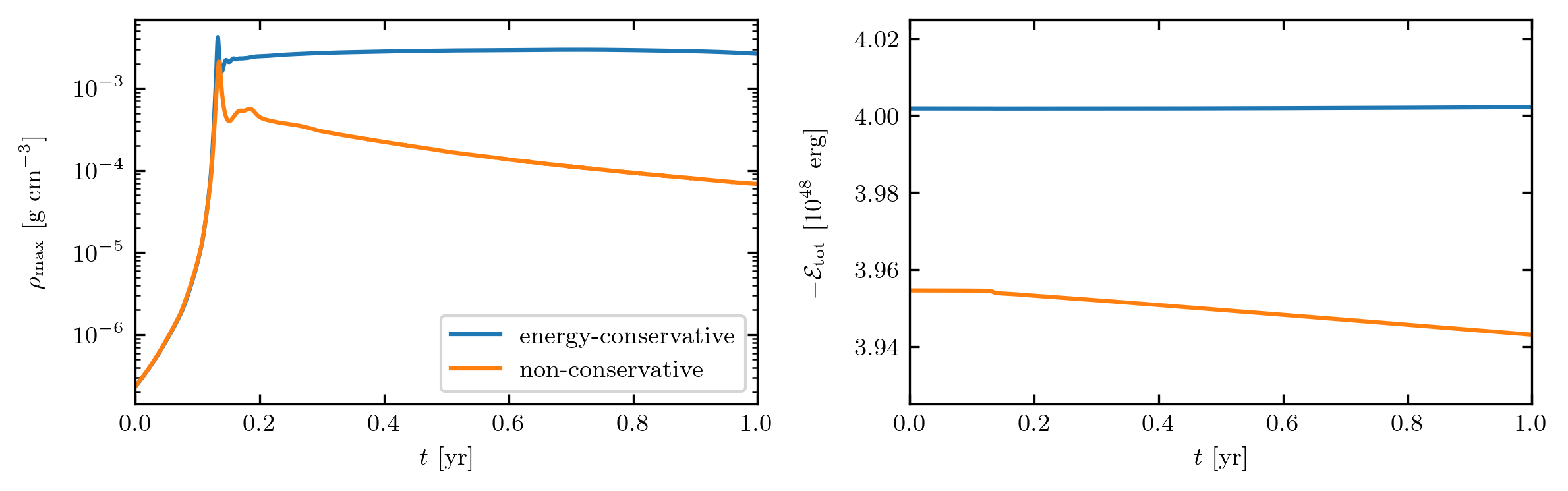}
    \caption{
        Comparison of spherical collapse simulations using the energy-conservative self-gravity scheme of \citet{Mullen_et_al_2021} and a conventional nonconservative scheme.
        The left panel shows the time evolution of the maximum density $\rho_\mr{max}$ in the computational domain, which corresponds to the central density of the protostar after its formation.
        The right panel presents the total energy $\mathcal{E}$ that should be conserved during the calculation, with its definition given in the text.
        Since $\mathcal{E}$ takes negative values due to the dominance of gravitational energy, we plot $-\mathcal{E}$ in the figure.
        The blue and orange lines indicate the results obtained with the energy-conservative and nonconservative schemes, respectively.
        The moment when the central density reaches $2.4\times10^{-7}~\mathrm{g~cm}^{-3}$ is defined as $t=0$.
    }
    \label{fig:GravSolver}
  \end{center}
\end{figure}
As described in Section~\ref{subsec:gravity_step}, we adopt the self-gravity scheme proposed by \citet{Mullen_et_al_2021}, which ensures total energy conservation with higher accuracy than the conventional approach and preserves the curl-free nature of the gravitational field, $\nabla \times \bm{g} = 0$.
Unlike conventional self-gravity updates, which treat gravity simply as source terms, their method evaluates the source terms in a way that is mathematically equivalent to flux divergences.
In this section, we compare the protostellar evolution obtained with this energy-conservative scheme to that from a conventional nonconservative one.
To this end, we simulate the formation and evolution of a protostar resulting from the collapse of a spherical cloud, incorporating primordial gas chemistry and modeling radiation with two frequency bins, following the approach described in Section~\ref{sec:ProsForm}.
As the initial condition, we adopt a critical isothermal Bonnor–Ebert sphere with a central density of $2.3\times10^{-15}~\mr{g~cm}^{-3}$ and a temperature of $4.8\times 10^3$~K \citep{Bonnor_1956}.
These values are taken from the evolutionary path of a collapsing primordial cloud under strong far-UV radiation, as given by the one-zone calculations of \citet{Omukai_2001}.
To induce collapse, the density profile is enhanced by a factor of $1.6$.
The radius of the Bonnor–Ebert sphere is $5.4\times10^3~\mr{au}$, while the length of the entire computational domain is $1.8\times10^4~\mr{au}$.
The gas is kept static outside the Bonnor–Ebert sphere, whereas radiation is solved throughout the domain and allowed to escape freely through the outer boundary.
The computational domain is adaptively refined to resolve the local Jeans length with at least eight cells.
We then follow the gravitational collapse and investigate the subsequent evolution of the protostar formed at the cloud center.
\par
The left panel of Figure~\ref{fig:GravSolver} shows the time evolution of the maximum density in the computational box.
Initially, this maximum density corresponds to the center of the collapsing cloud, and after the formation of a protostar, it reflects the central density of the protostar itself.
The blue and orange lines indicate the results obtained with the energy-conservative and nonconservative self-gravity schemes, respectively.
In both cases, the gas density initially increases due to gravitational contraction. When the density reaches $\rho \sim 10^{-3}~\mathrm{g~cm}^{-3}$, the core becomes optically thick, the radiative cooling becomes inefficient, and the cloud collapse halts, marking the formation of the protostar. 
Up to this point, the evolution proceeds nearly identically in both simulations.
However, significant differences emerge in the subsequent evolution.
With the energy-conservative scheme, the central density remains nearly constant over time, whereas, in the nonconservative case, it gradually decreases.
At $t=1$~yr, the central densities differ by more than an order of magnitude.
This divergence arises because the nonconservative scheme introduces a gradual increase in the total energy of the system, which artificially raises the entropy in the protostellar interior.
The right panel of Figure~\ref{fig:GravSolver} shows the time evolution of the total energy $\mathcal{E}_\mathrm{tot}$.
Here, we calculate the total energy $\mathcal{E}_\mr{tot}$ as
\begin{eqnarray}
    \mathcal{E}_\mr{tot} = \sum_{i} \Delta V^i (E^i_\mr{gas}+E^{\mr{con},i}_\mr{rad}+E^i_\mr{grav}+\Delta E^i_\mr{chem}) + \Delta \mathcal{E}_\mr{rad,esc} ,
\end{eqnarray}
where superscript $i$ is the cell index in the computational domain, $\Delta V$ is the cell volume, and $E_\mr{grav}=-\rho\phi/2$ is the gravitational energy.
The fifth term $\Delta E_\mathrm{chem}$ represents the chemical potential energy relative to a fully neutral gas, which is given by
\begin{eqnarray}
    \Delta E_\mr{chem} = n_\chH (\chi_{\chH} y(\chHp) - \chi_{\chHt} y(\chHt) - \chi_{\chHm} y(\chHm) ),
\end{eqnarray}
where $y(\mathrm{A})$ is the number fraction of species $A$ relative to hydrogen nuclei.
Here, $\chi_\chH = 13.6$~eV is the ionization energy of hydrogen, $\chi_{\chHt} = 4.48$~eV is the dissociation energy of molecular hydrogen, and $\chi_{\chHm} = 0.755$~eV is the detachment energy of the negative hydrogen ion.
The last term $\Delta \mathcal{E}_\mr{rad,esc}$ is the cumulative radiation energy that has escaped from the computational box, which is obtained by
\begin{eqnarray}
    \Delta \mathcal{E}_\mr{rad,esc} = \sum_{n} \Delta t^n \int_{\partial V} (\bm{F}^n_\mr{S}+\bm{F}^n_\mr{num}+\bmFofNonRSLAn_\mr{diff}) \cdot d\bm{S} ,
\end{eqnarray}
where the superscript $n$ denotes the time step index, $\Delta t^n$ is the value of the $n$ th time step, $d\bm{S}$ is the area element, and the integration is performed over the boundary $\partial V$ of the computational domain.
Although the total energy $\mathcal{E}_\mr{tot}$ is not strictly conserved because the gas outside the Bonnor–Ebert sphere is artificially kept static by the boundary condition, the central high-density region dominates the energy budget. Therefore, the influence of the boundary treatment is negligible, and the total energy is expected to remain constant.
In the present setting, the gravitational energy dominates, and thus, $\mathcal{E}_\mathrm{tot}$ takes negative values.
As seen in the right panel of Figure~\ref{fig:GravSolver}, in the conservative scheme, $\mathcal{E}_\mathrm{tot}$ remains nearly constant with the relative difference between $t=0$ and $t=1$ being only $10^{-4}$.
In contrast, in the nonconservative scheme, $\mathcal{E}_\mathrm{tot}$ is larger than in the conservative case throughout the evolution, and after protostar formation, it gradually increases with time, i.e., its absolute value decreases, indicating a net gain of energy.
Such an artificial increase in energy and the associated density decrease has also been reported by \citet{Mullen_et_al_2021} in hydrostatic systems without accretion.
These results highlight the importance of employing a conservative self-gravity scheme to accurately follow the evolution of self-gravitating systems such as protostars.

\end{document}